\documentclass{aa}
\usepackage{graphicx, epsfig, fancyhdr, rotating, amsmath, epsf, txfonts, natbib, epstopdf, multirow}
\usepackage{subfigure, tabularx, appendix, dcolumn, threeparttable, longtable, makecell,appendix}
\usepackage{psfig}
\usepackage[colorlinks=true,allcolors=blue]{hyperref}
\usepackage[normalem]{ulem}
\usepackage{hyperref}
\newcount\inclfigs \inclfigs=1

\def\kms{\hbox{km$\;$s$^{-1}$}} 
\def\arcsec{\hbox{$^{\prime\prime}$}}

\def\aahh{\hbox{AIA~193}}
\def\aah{\hbox{AIA~171}}
\def\aac{\hbox{AIA~304}}

\def\aahhh{\hbox{AIA~94}}
\def\mg{\hbox{Mg~{\sc ii} k~2796}}
\def\c{\hbox{C~{\sc ii}~1335}}
\def\si{\hbox{Si~{\sc iv}~1394}}
\def\sj{\hbox{SJI~1400}}

\begin{document}

\title{Eruptions from coronal bright points: A spectroscopic view by IRIS of a mini-filament eruption, QSL reconnection, and reconnection-driven outflows}
\titlerunning{Eruptions from coronal bright points: A spectroscopic view by IRIS}

\author{Maria~S. Madjarska\inst{1}, Duncan~H. Mackay\inst{2}, Klaus Galsgaard\inst{2}, Thomas Wiegelmann\inst{1}, Haixia Xie\inst{3} }

\offprints{madjarska@mps.mpg.de}
\institute{
Max Planck Institute for Solar System Research, Justus-von-Liebig-Weg 3, 37077, G\"ottingen, Germany
\and
School of Mathematics and Statistics, University of St Andrews, North Haugh, St Andrews, KY16 9SS, Scotland, UK
\and
Fundamental Teaching Department, Shandong Jiaotong University, 264209, Weihai, Shangong, China}

\date{Received date, accepted date}

\abstract
{The present study investigates a mini-filament eruption associated with cancelling magnetic fluxes. The eruption originates from a small-scale loop complex commonly known as a Coronal Bright Point (CBP). The event is uniquely recorded in both the imaging and spectroscopic data taken with the Interface Region Imaging Spectrograph (IRIS).}
{This investigation aims to gain a better understanding of the physical processes driving these ubiquitous small-scale eruptions.}
{We analyse IRIS spectroscopic and slit-jaw imaging observations as well as images taken in the extreme-ultraviolet channels of the Atmospheric Imaging Assembly (AIA), and line-of-sight magnetic-field data from the Helioseismic Magnetic Imager (HMI) on board the Solar Dynamics Observatory. As observations can only indicate the possible physical processes at play, we also employ a non-linear force-free field (NLFFF) relaxation approach based on the HMI magnetogram time series. This allows us to further investigate the evolution of the magnetic-field structures involved in the eruption process. }
{We identify a strong small-scale brightening as a micro-flare in a CBP, recorded in emission from chromospheric to flaring plasmas. The mini-eruption manifests with the ejection of hot (CBP loops) and cool (mini-filament) plasma recorded in both the imaging and spectroscopic data. The micro-flare is preceded by the appearance of an elongated bright feature in the IRIS slit-jaw 1400~\AA\ images located above the polarity inversion line.  The micro-flare starts with an IRIS pixel size brightening and propagates bi-directionally along the elongated feature. We detect in both the spectral and imaging IRIS data and AIA data, strong flows along and at the edges of the elongated feature which we believe represent reconnection outflows. Both edges  of the elongated feature that wrap around the edges of the erupting MF evolve into a J-type shape creating a sigmoid appearance. A quasi-separatrix layer (QSL) is identified in the vicinity of the polarity inversion line by computing the squashing factor $Q$ in  different horizontal planes of the NLFFF model.}
{This CBP spectro-imaging study provides further evidence that CBPs represent downscaled active regions and as such may have a significant contribution to the mass and energy balance of the solar atmosphere. They are the sources of all range of typical active-region features including, magnetic reconnection along QSLs, (mini-)filament eruptions, \mbox{(micro-)flaring}, reconnection outflows, etc. The QSL reconnection site has the same spectral appearance as the so-called explosive events identified by strong blue- and red-shifted emission, thus answering a long outstanding question about the true nature of this spectral phenomenon.}

\authorrunning{Madjarska et al.}

\maketitle

\section{Introduction}
\label{intro}


\begin{figure*}[!ht]
\center
\centerline{\includegraphics[scale=0.4]{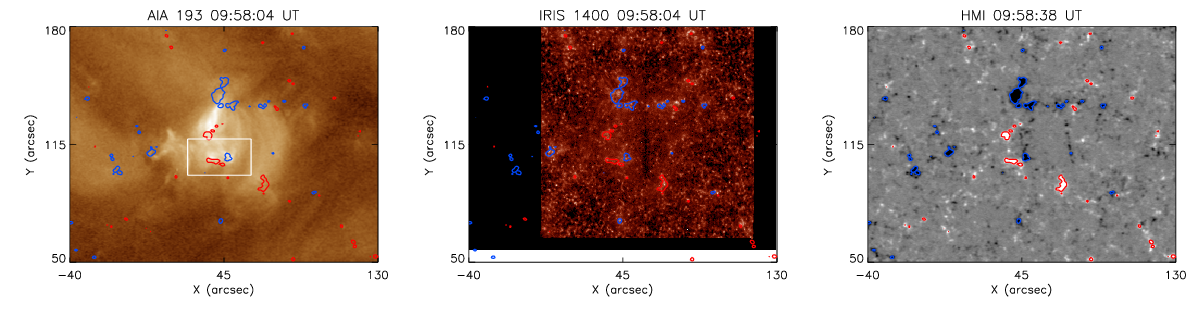}} 
\caption{Small-scale loop system known as a coronal bright point. {\it From left to right:} \aahh~\AA, IRIS~SJI~1400, and HMI magnetogram images with the CBP in the centre of the field-of-view (FOV). The red and blue contours trace magnetic fluxes at $\pm$50~G. The HMI magnetograms are saturated at $\pm$50~G. The white-line square in the first panel outlines the FOV shown in Fig.~\ref{fig2}. }
\label{fig1}
\end{figure*}

In recent years we have reported three studies on eruptions from small-scale loop systems known as coronal bright points (CBPs) by \citet[][hereafter Paper~I]{2018A&A...619A..55M}, \citet[hereafter Paper~II]{2019A&A...623A..78G}, and \citet[hereafter Paper~III]{2020A&A...643A..19M}. These eruptions have been named jets in Extreme-ultraviolet (EUV) and X-ray data \citep[e.g.][and references therein]{2020A&A...643A..19M} when detected in coronal holes and mini Coronal Mass Ejections \citep[mini-CMEs][]{2009A&A...495..319I,2018A&A...619A..55M} while observed in the quiet Sun. The term mini-CME is used to describe the observation of small-scale eruptions that do not evolve as collimated flows, i.e. as jets typically ejected from CBPs in coronal holes. Rather they develop as expanding bubbles due to the closed magnetic-field structure of the quiet Sun. Whether these eruptions are fully confined or if some of the ejected material reaches into the interplanetary space remains an open question. We will refer hereafter to this phenomenon simply as mini-eruptions. The CBPs are composed of a set of small-scale coronal loops seen with enhanced emission in
EUV and X-rays linking photospheric magnetic-flux
concentrations of opposite polarity \citep[for review see][]{2019LRSP...16....2M}.1

We will briefly review the main results on the eruptions from Papers~I, II, and III. Paper~I shows that 76\%\ of the analysed CBPs (31 out of 42) were the source region of one or more eruptions. The eruptions generally occurred during the late stage of the life of the CBPs when magnetic-flux convergence and cancellation are typically occurring. The CBP eruptions commonly evolve with the ejection of cool and hot plasma. The cool plasma appears in absorption in the coronal imager channels and is related to mini-filaments (MFs) while the hot plasma represents the expelled CBP loops and the overlying corona. Micro-flares always appear at the CBPs polarity inversion line (PIL) during CBP eruptions. We identified dimmings caused by MF's `dark' appearance and reduced emission due to density depletion. In Paper~II we proceeded with non-potential time-dependent modeling of the CBP magnetic field. 
A non-linear force-free field (NLFFF) relaxation approach using a time series of Helioseismic and Magnetic Imager (HMI) line-of-sight magnetograms was employed. Details on the methodology can be found in Section 4 in Paper~II and Appendix~A in Paper~III. 
The results revealed that most of the CBP eruptions in Paper~II are related to the formation of twisted magnetic flux, i.e. flux ropes, that form at the location of the eruptions. The nature of the micro-flares remained unclear and was explored in Paper~III where a large CBP located in an equatorial coronal hole was investigated. In this study, it was found that the formation and destabilisation of mini-filaments cause the occurrence of a EUV/X-ray jet. The mini-filament formed along the polarity inversion line of the CBPs 3--4 hrs before its eruption seen in GONG (Global Oscillation Network Group) H$\alpha$ images. In this case, the bipolar magnetic flux was not seen to converge or cancel during the time leading to the MF destabilisation and eruption. Most importantly the eruption-associated micro-flare was identified with three flare kernels seen in H$\alpha$ images appearing shortly after the MF lift-off. The derived magnetic NLFFF model reproduced features explaining various observational characteristics related to the evolution of the CBP and the eruption of the MF. 

In the present paper, we investigate a mini-filament eruption from a CBP associated with a micro-flare and cancelling magnetic fluxes. The IRIS spectral and imaging observations capture the energy release site associated with the mini-eruption, revealing crucial information on how magnetic reconnection and associated phenomena proceed during cancelling bipolar features detected in magnetic-field observations. To shed more light on the configuration and evolution of the coronal magnetic field associated with the observed complex phenomenon and the physical processes that take place, we also employ a non-linear force-free field (NLFFF) relaxation modeling, followed by squashing factor computation. Section~\ref{obs} reports on the observations from IRIS, AIA, and HMI. The modeling methodology is given in full detail in papers~II and III. The observational and modelling results are presented in Section~\ref{res} and the discussion in Section~\ref{disc}. The summary and conclusions are given in Section~\ref{concl}.


\begin{figure*}[!ht]
\center
\centerline{\includegraphics[scale=0.5]{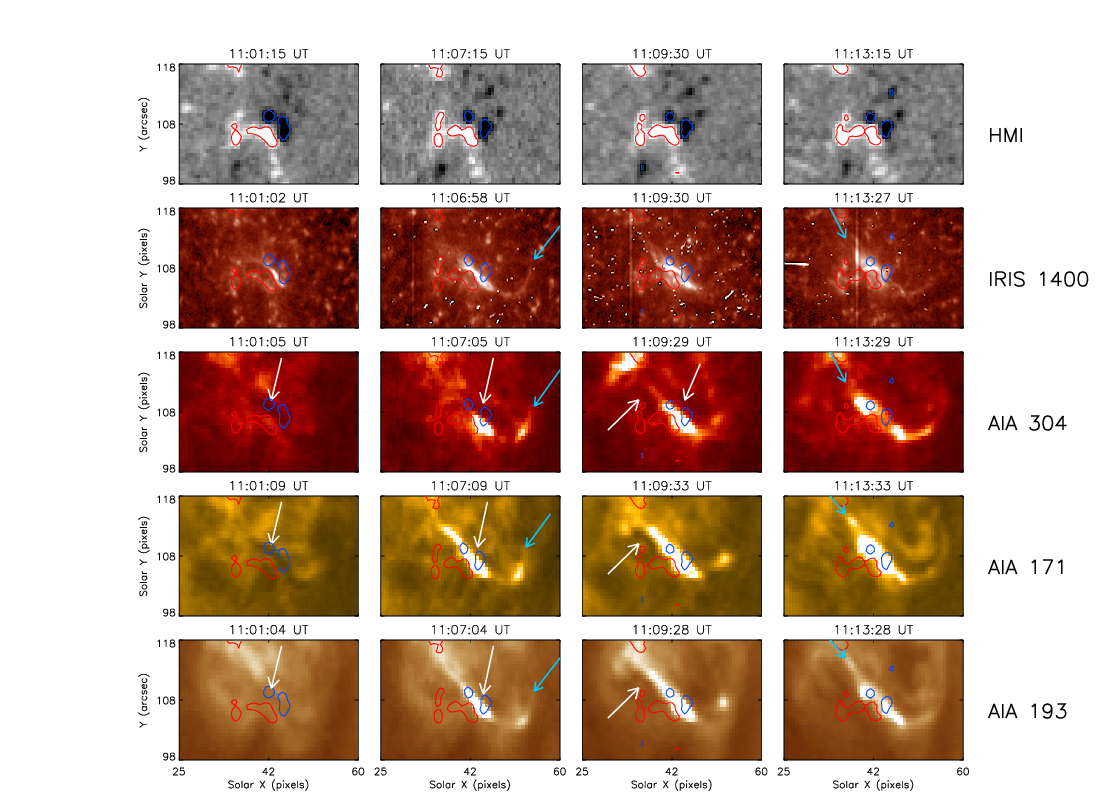}} 
\caption{{\it From top to bottom: } HMI magnetograms, IRIS SJI~1400, \aac, \aah, and \aahh\ images showing the evolution of the eruption.
The white arrows point at the mini-filament seen in absorption, i.e. dark feature. The red and blue contours trace magnetic fluxes at $\pm$50~G. The cyan arrows indicate the hot plasma up-flows seen as bright extensions from the reconnection site.}
\label{fig2}
\end{figure*}


\begin{figure*}[!ht]
\center
\centerline{\includegraphics[scale=0.5]{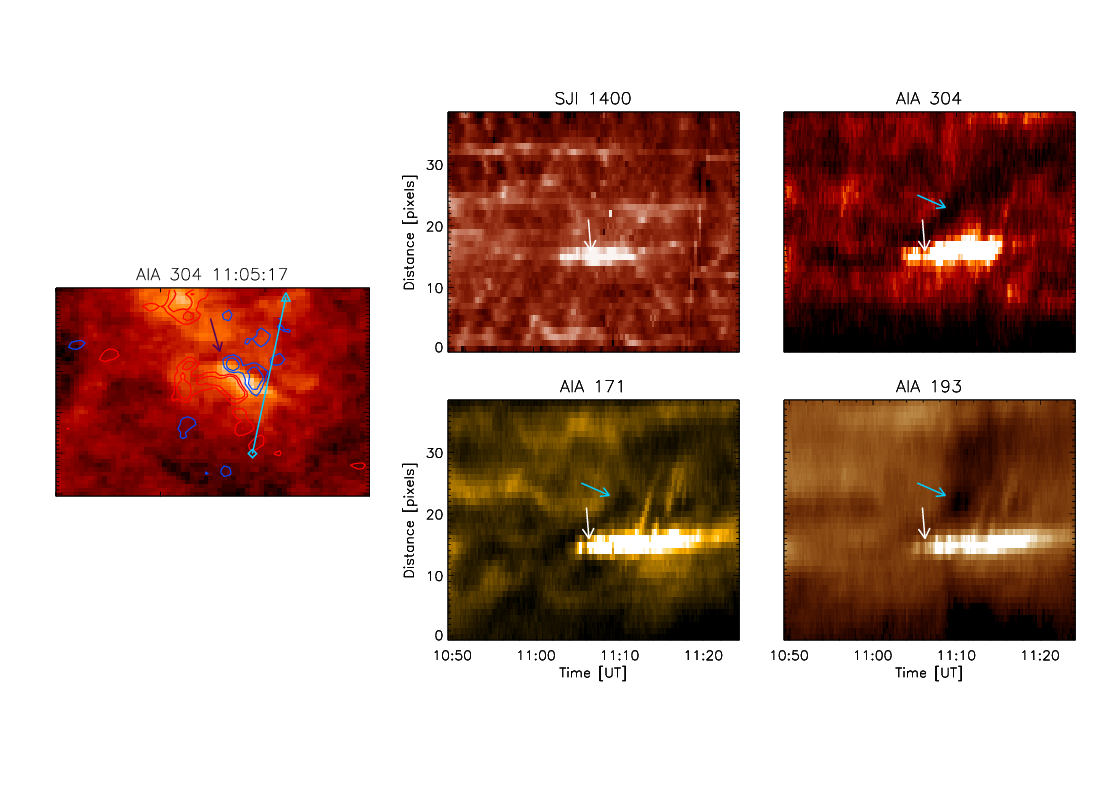}} 
\vspace{-1cm}
\caption{{\it From left to right:} The \aac\ image during the mini-filament appearance as a dark elongated feature. A cyan line marks the slice from which the time-slice images in the second and third columns are produced. The square indicates the bottom and the triangle, the top of the slice images. The black arrow points at the mini-filament. The red and blue contours trace magnetic fluxes at $\pm$25~G and $\pm$50~G. {\it Top middle:} IRIS SJI ~1400 time slice image. The white arrow points at the micro-flare (the same for the AIA images). {\it Bottom middle:} \aah\ time-slice image. {\it Top right:} \aac\ time slice image, and {\it Bottom right:} \aahh\ slice image. The cyan arrows point at the dimming region.  The IRIS pixels are rescaled to the AIA pixel size.}
\label{fig3}
\end{figure*}

\section{Observational data}
\label{obs}

On 2017 April 5 during a dedicated observing campaign of CBPs, the Interface Region Imaging Spectrograph (IRIS) \citep{2014SoPh..289.2733D} took observations of a CBP located not far from the disc centre (see Fig.~\ref{fig1}, first panel). The observations include slit-jaw images taken in the 1400~\AA\ (hereafter \sj) with a field-of-view (FOV) of 119\arcsec\ $\times$ 119\arcsec\ and 17~s cadence and a pixel size of 0.17\arcsec\ $\times$ 0.17\arcsec. The IRIS spectral data were taken in a rastering mode with a 17 s cadence, and a 15 s exposure time.
The width of the IRIS slit is 0.35\arcsec\ while the pixel size in solar Y is 0.17\arcsec. Seven spectral windows were transferred to the ground. In this study, we analyse the 
Si~{\sc iv} 1393.75~\AA\ line that has a maximum formation temperature of 63\,000 K as well as the optically thick C~{\sc ii} 1334.53~\AA\ and Mg~{\sc ii} k 2796.35~\AA\ lines that have more complex formation. In general, the Mg~{\sc ii} lines sample the mid-chromosphere \citep[for details see][]{2013ApJ...772...90L}, while C~{\sc ii} the upper chromosphere and low solar transition region \citep{2015ApJ...811...81R}. The C~{\sc ii} lines typically form above the Mg~{\sc ii} lines. Further information on the Mg~{\sc ii} lines can be found at \url{https://iris.lmsal.com/itn39/Mg_diagnostics.html} and on the C~{\sc ii} lines at \url{https://iris.lmsal.com/itn39/CII_diagnostics.html}.

We also analysed imaging data from AIA \citep{2012SoPh..275...17L} on board SDO \citep{2012SoPh..275....3P} taken in the EUV 94~\AA\ (hereafter \aahhh), 193~\AA\ (hereafter \aahh), 171~\AA\ (hereafter \aah), and 304~\AA\ (hereafter \aac) channels. A description of the response of the three EUV channels is given in Paper~I and III, as well as in \cite{2021A&A...646A.107M}. The AIA EUV data have a 12~s cadence and 0.6\arcsec\ $\times$ 0.6\arcsec\ pixel size. We used HMI line-of-sight magnetograms \citep{2012SoPh..275..207S} that have both, a 45~s and a 720~s cadence. The HMI data have a 0.5\arcsec\ $\times$ 0.5\arcsec\ pixel size but were rescaled to the AIA pixel size of 0.6\arcsec\ using the hmi\_prep.pro procedure. The AIA 1600~\AA\ images were used to align the HMI data with those from the EUV channels and IRIS with a precision of the alignment at $\sim$1\arcsec. All images were
de-rotated to 10:00~UT on April 5.

\begin{table*}[h!]
\caption{Timeline of the eruption.}
\label{table1}
\centering
\begin{tabular}{ c l }
\hline
Time (UT) & Event\\
\hline
04:00 & the bipole forms\\
05:30 & interconnecting loops seen in \aac\\
06:00& cancellation starts \\
09:45& the magnetic flux of the bipole starts to decrease\\
11:00& bright elongated feature along the PIL in \sj\\
11:00& MF starts to appear in the AIA EUV channels\\
11:00:45 & first brightening appears in \sj\\
11:02:41 & first brightening appears in \aac\\
$\sim$11:03 & first brightening appears in AIA~193 and AIA 171\\
$\sim$11:04 & first brightening appears in AIA~94\\
$\sim$11:06& quick coronal intensity depletion\\
$\sim$11:06& MF lifts-off\\
$\sim$11:06& outflows appear\\
\hline
\end{tabular}
\end{table*}

\section{Results}
\label{res}

The IRIS spectroscopic and imaging observations are taken in the far-ultraviolet (FUV) and near-ultraviolet (NUV) wavelength ranges, and provide information about the physical properties and dynamics of the chromospheric and transition-region plasmas in the quiet Sun of the solar atmosphere. These combined with AIA transition-region and coronal imaging observations give a complete overview of the emission from plasmas with a wide range of temperatures. Such an assembly of data can sample emission from the upper photosphere in the AIA~1600 channel (together with some transition region emission from C~{\sc iv} lines), the mid and upper chromosphere, in e.g. Mg~{\sc ii} and C~{\sc ii}, up to emission from a flaring corona in the AIA~94~\AA\ channel with a response of up to 7~MK \citep[Fe~{\sc xviii}, for details, see][]{2010A&A...521A..21O, 2013A&A...558A..73D}.

As we mentioned in Section~\ref{obs}, the observations analysed here were taken during a dedicated CBP observing campaign when CBPs close to the disc centre were targeted. As the IRIS planning requires a target selection a day in advance, it means that the observed CBPs would often be in the late stage of their lifetime given the fact that the CBP average lifetime is 24~h \citep[see section 3.2 in][]{2019LRSP...16....2M}.

The first quick look at the AIA and IRIS data taken on 2017 April 5 indicated that during the second of the two IRIS rasters both hot and cool plasma ejections occurred. The
small-scale plasma ejection of cool plasma is associated with a mini-filament while the hot plasma is
linked to the bright point coronal loops and micro-flare (McF) heated plasmas taking place at the lower (southern) part of the targeted CBP. The source region of the eruption is shown in Fig.~\ref{fig1} in the centre of the area outlined with a white square. The evolution of the phenomenon can be followed in the animation provided in  Fig.~\ref{app1} and in Fig.~\ref{fig2}, where near co-temporal multi-instrument images are shown. To the best of our knowledge, this is the first report of a mini-eruption from a small-scale loop system (a CBP) in the quiet Sun involving a mini-filament that has been detected in the IRIS imaging (slit-jaw) but also in the spectral data, i.e. the spectrometer slit is scanning the site while the eruption is ongoing.

At first glance, the evolution of the eruption in the high-resolution \sj\ looked very intriguing as certain details never detected before in AIA images due to resolution limitations were seen. The spectral observations revealed even more exciting features as stronger blue- and weaker red-shifted emission were on display in all spectral lines while the slit was crossing the eruption site. This type of spectral appearance has been usually attributed to the so-called `explosive events' (EEs) \citep[e.g.][]{1983ApJ...272..329B,1984ApJ...281..870D} that are interpreted as the observational signature of bi-directional flows resulting from magnetic reconnection \citep[e.g.][]{1991JGR....96.9399D}. We come back on this in Section~\ref{disc}. The co-temporal imaging information suggested a highly dynamically evolving {eruptive phenomenon} associated with  a mini-filament eruption and a micro-flare \citep{2009A&A...495..319I}. To inspect this we then analysed all available data including magnetic and intensity imaging (Section~\ref{er_im}) and spectral (Section~\ref{er_sp}) observations. As observations can only provide an inkling of the possible physical processes in play, we
also performed data-driven simulations (see Section~\ref{er_mod}) to obtain a deeper and more accurate insight into the physical processes involved in the observed phenomenon.

\begin{figure*}[!ht]
\vspace{-2cm}
\center
\centerline{\includegraphics[scale=0.45]{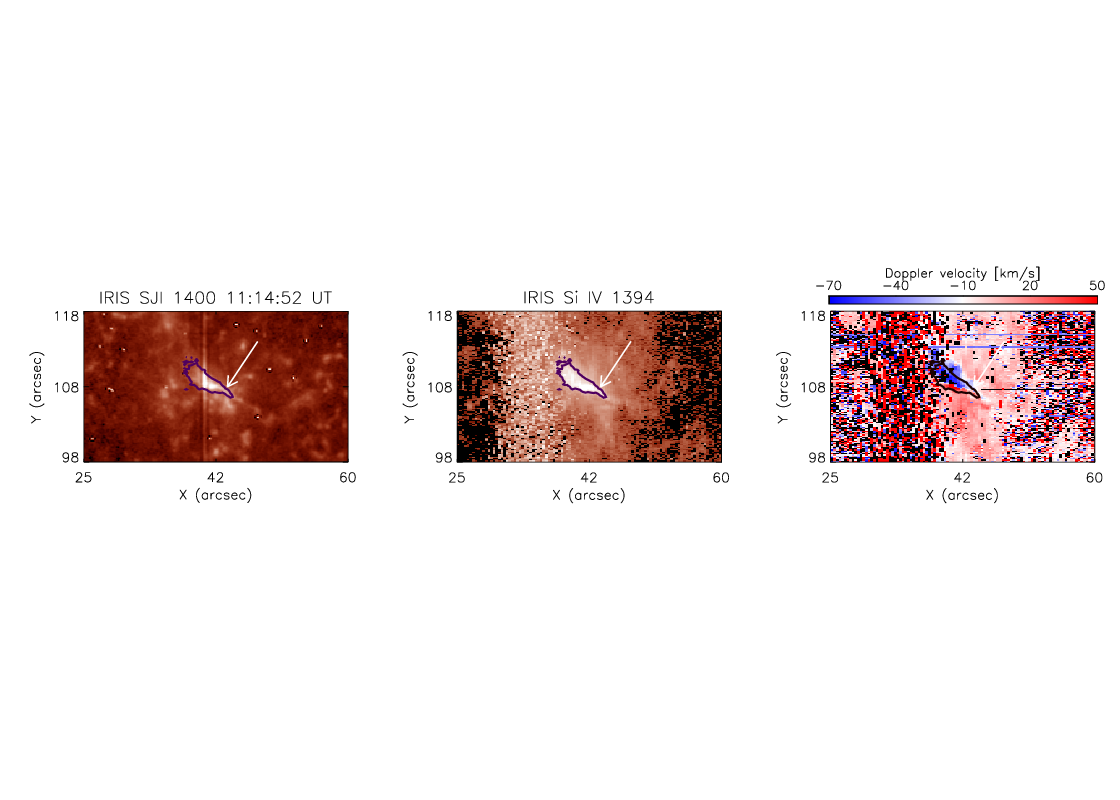}} 
\vspace{-4cm}
\caption{{\it From left to right:} IRIS SJI 1400 image, Si~{\sc iv} 1394 raster intensity and Doppler velocity images. The contour outlines the micro-flare region in the \si\ raster image that is also overplotted on the \sj\ and Doppler-velocity images. The white arrow points at the reconnection site.}
\label{fig5}
\end{figure*}

\subsection{The eruption in the imaging data} 
 
\label{er_im}

The source region of the eruption is a bipole. Its location is shown in the centre of the top row of the images in Fig.~\ref{fig2}. The bipole opposite polarity fluxes arise from the convergence of the negative flux of a newly emerging bipole with a pre-existing positive flux concentration. The negative flux increases in strength while approaching the positive flux which remains relatively stationary. The region forms as a discernible bipole around 04:00~UT on April 5. Interconnecting loops become distinguishable in the \aac\ channel almost an hour and a half later, at $\sim$05:30~UT. A few minutes later interconnecting loops also appear in the high-temperature \aah\ and 193 channels. The cancellation between the two opposite polarities starts at $\sim$06:00~UT. CBPs are known to be formed by loops at different temperatures, where hotter larger loops are reported to overlie cooler lower ones \citep{2019LRSP...16....2M}. The bipole total magnetic flux starts to decrease around 09:45~UT and falls sharply until at least 11:28~UT with a total absolute flux decrease of about 24\%, i.e. the measurable cancellation has started earlier than the eruption analysed here. The magnetic flux decrease is about 13\%\ between the detectable beginning 10:59~UT and the end of the eruption at 11:28~UT. The timeline of the events can be followed in Table~\ref{table1}.

The transition-region and coronal activity above the bipole location begin around 09:30~UT with two mini-eruptions very similar to the one investigated in detail here. They are however weaker, especially in the AIA imaging data. The first eruptive event occurred during the time period between 09:40~UT and 09:55~UT (scanned by the IRIS slit) and the second between 10:15~UT and 10:25~UT (not scanned by the IRIS slit). Repetitive eruptions (homologous) from the same cancelling bipole are a very common feature and have been shown and discussed in detail in 
\citet{2018A&A...619A..55M} and \citet{2019A&A...623A..78G}.

The third in the series of eruptions is studied here in full detail. It begins with the appearance of a bright elongated feature (the same is seen during the first eruption) that can only be seen in the IRIS SJIs starting at 11:00~UT  (see the animation in Fig.~\ref{app1}). The feature extends above the polarity inversion line separating the two opposite polarities. It has a cross-section of approximately two IRIS pixels or 0.34\arcsec\ ($<$250~km) which is one of the possible explanations for why this feature is only distinguishable in the IRIS data and not in the AIA data that have a pixel size of 0.6\arcsec.

The dynamic event starts with a pixel size brightening at the elongated feature seen in the \sj\ at 11:00:45~UT that expands bi-directionally along the elongated feature. 
This type of brightening is considered to be the observational signature of a micro-flare. It first appears in AIA 304 at 11:02:41~UT. In the coronal channels AIA~193 and AIA 171 it is seen at 11:03:16~UT and 11:03:09~UT, respectively, which can be considered co-temporal given the fact that the two channels do not operate simultaneously. To determine with more certainty that the observed brightening is related to impulsive heating at a few million degrees (flare-type), we investigated the emission in the \aahhh\ channel. This channel contains the Fe~{\sc xviii} 93.932~\AA\ (maximum formation temperature of 7.08~MK) but also emission from spectral lines with lower formation temperatures like Fe~{\sc x, xii}, etc. Detailed information and analysis on this can be found in \citet{2013A&A...558A..73D}. To remove the `contamination' by the cooler emission we used the emission from the event in the AIA~211 and 171 channels. We subtracted this `cooler emission' by applying the expression given by \citet{2013A&A...558A..73D}. The resulting response in the AIA~94 can be followed in Fig.~\ref{app1} with the channel named AIA Fe~{\sc xviii}. The plasma at the initial micro-flare location first reached very high temperatures at 11:04~UT. The micro-flare signature appears below the rising mini-filament as has been observed in our earlier cases of mini-eruptions \citep[for more details and additional examples see][]{2018A&A...619A..55M}.  Because of the obscuration 
caused by the rising MF \citep[for more details see][]{2018A&A...619A..55M,2020A&A...643A..19M}, a lightcurve (not presented here) shows a clear presence of hot plasma only between $\sim$11:06~UT and $\sim$11:10~UT. 

The inspection of the general evolution of the CBP (see the first panel of Fig.~\ref{fig1}) indicated that the CBP as seen in \aahh\ on April 5 is part of a larger loop system with a configuration and evolution that appears intricate due to a complex distribution of magnetic flux at a photospheric level. In other words, the loops forming the CBP are not rooted in a simple bipolar region. This type of CBPs has been described already in our previous studies \citep[see e.g.][]{2018A&A...619A..55M}, thus we will not go into further details here.

A mini-filament starts to appear in the AIA EUV images around 11:00~UT (see the animation in Fig.~\ref{app1}) and it becomes visible as an elongated absorption (dark) feature shortly before its eruption. The white arrows in the second and third columns in Fig.~\ref{fig2} point at the MF. An 
enlarged view is also given in the first panel of Fig.~\ref{fig3}, left panel. The MF lies along the polarity inversion line (PIL) separating the bipole mentioned above. The appearance of the MF is not exactly between the two polarities which can be explained by the asymmetry of the magnetic configuration (for more details see section~\ref{er_mod}).
The MF has an apparent minimum
length of 8.7~Mm, which is below the average size of mini-filaments at 14.5~Mm (15\arcsec) \citep{1986NASCP2442..369H}. As the filament length is
determined in the coronal imaging data and not in the chromospheric line \citep[e.g. H$\alpha$, see discussion on the visibility in][]{2020A&A...643A..19M}, the true length could be larger than the estimated one. After a slow rise between roughly 11:01~UT and 11:06~UT, the MF rapidly lifts off which can be followed in the time-slice images in Fig.~\ref{fig3}  and in Fig.~\ref{app1}. The \aac\ gives the best view of the MF eruption. \citet{2010A&A...517L...7I} using data from the twin EUVI/SECCHI telescopes on the STEREO spacecraft, i.e. a combination of on-disk and off-limb data, have found that the removal of the overlying corona, seen as coronal emission depletion,  occurs earlier than the rise of the MF. The dimming region is caused by reduced coronal density and the presence of the erupting mini-filament seen in absorption. 
For the case of the eruption studied here, AIA 193 shows better the dimming region than AIA 171. The dimming is least visible in AIA 304 apart from the footpoint of the erupting mini-filament. The AIA 171 and especially the AIA 193 channel show a more extended dimming region earlier than the mini-filament eruption which could be due to the removal of the overlying corona as mentioned above. However, we cannot be certain because of the strong background emission. It should be noted that a brighter feature runs across the dimming region which is best seen in AIA~171. It is possibly related to some of the CBPs expanding loops as seen in many similar cases by \cite{2018A&A...619A..55M}. 

The most probable explanation for the different appearance of the dimming region in the different AIA EUV channels is related to  the intensity of registered background emission and its temperature, and the level of coronal depletion at a given temperature during the eruption. The AIA 304 channel is dominated by transition-region emission. The AIA 171 channel also contains strong background emission at low temperatures (for details on the response of the AIA EUV channels see section~3 in Paper~I). Thus generally the dimming will have  a different appearance in different EUV channels. Examples of various appearances of dimming regions can be seen in \citet{2018A&A...619A..55M} and \citet{2020A&A...643A..19M}.  The MF is not visible in the SJI~1400, only the micro-flare. Generally, filaments/prominences emit at transition-region temperatures which comes from the so-called prominence-corona transition region \citep[e.g.][]{2004SoPh..223...95C,1999ESASP.446..467M} that separates the cool plasma of the filament core from the hot corona and can be distinguished only when observed above the limb. On disc, when seen as filaments, the emission is far too weak in comparison to the background chromospheric and transition-region emission of the quiet Sun, and that is why the erupting filament cannot be distinguished in the IRIS SJ~1400 channel.

As seen in the SJI~1400, a brightening associated with a micro-flare, as mentioned above, spreads bi-directionally along the elongated feature until $\sim$11:03~UT when both its edges rapidly (in 17~s) expand in length forming J-shaped edges (for details see the animation in Fig.~\ref{app1}).  The J-shaped edges give the whole feature the appearance of a sigmoid. The link between filament eruptions and CMEs in active regions and sigmoids' appearance in EUV and X-rays, has long been established \citep[e.g.][]{1996ApJ...464L.199R,2011A&A...526A...2G,2017ApJ...834...42J, 2014ApJ...782...71G}. The sigmoid-like feature coincides with the rise of the MF with the J-shaped edges wrapping around the two legs of the rising MF. This is a strong indication of the presence of a QSL. The spreading of the bright emission seen in all EUV channels has already been observed in larger CBP eruptions by \cite{2020A&A...643A..19M}. It has been associated with the micro-flare ribbons that are seen here as one single feature (due to the event's small size and limited resolution) rather than two bright lanes as typically observed in large solar flares \citep{2011SSRv..159...19F,2014ApJ...788...60J,2017LRSP...14....2B}. The micro-flare ribbons are the actual footprints of a current layer that has formed along
the QSL, extending from coronal to photospheric heights. Figure 7 in \citet{2014ApJ...788...60J} nicely illustrates this.
 
  Plasma outflows  are observed at two edges of the elongated feature best seen in the IRIS SJ images (for details see the animation in Fig.~\ref{app1}). The physical nature of these outflows is discussed in section~\ref{disc}. At 11:06~UT the west edge curves upwards (i.e. J-shaped) which coincides with the lift-off of the filament as mentioned above (follow the animation in the AIA 304 and the SJ 1400 images). At 11:15~UT this extension and associated  outflows are no longer visible in SJI~1400 but they remain present in the AIA~304 and coronal channels. The brightening along the whole structure starts to decrease after 11:10~UT initially from the west edge, close to where the initial brightening occurred while the eastern part continues being active. The elongated brightening event lasts at transition region temperatures until 11:31~UT (SJI~1400), while at coronal (AIA~171 and 193) it can be seen for at least another 10 min. 
 
\subsection{The eruption in spectroscopic data}
\label{er_sp} 

\begin{figure*}[!ht]
\hspace{1cm}
\centerline{\includegraphics[scale=0.5]{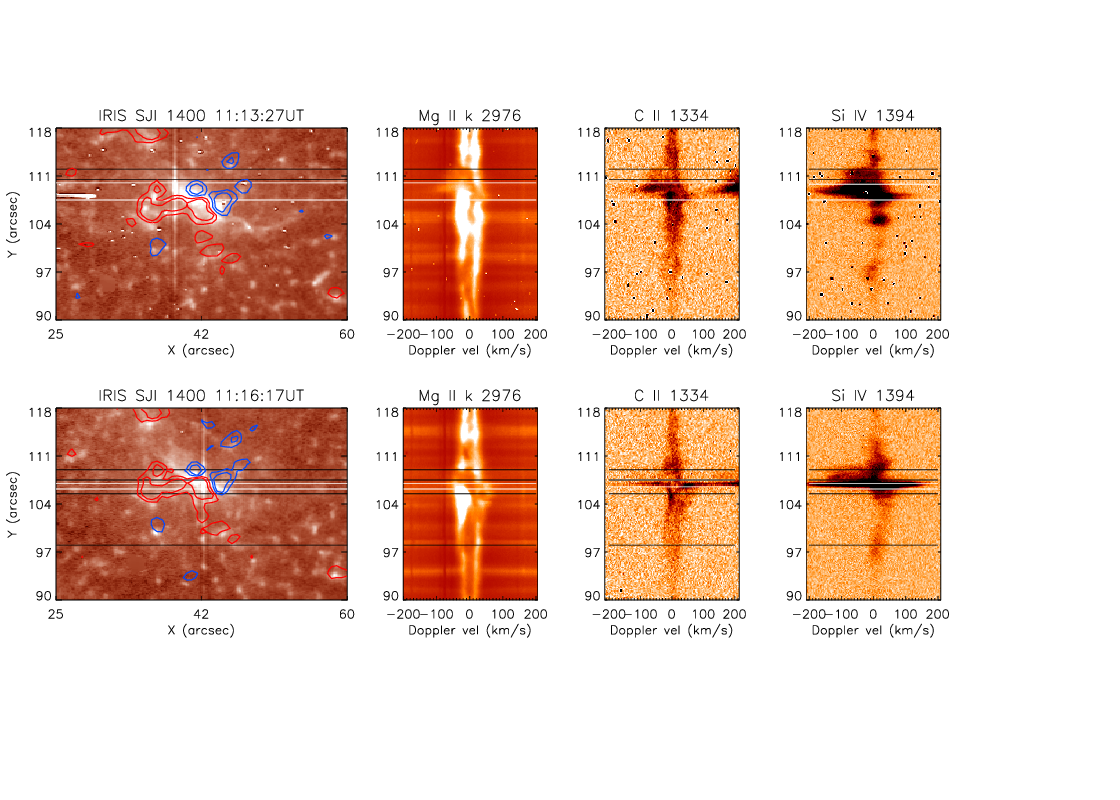}} 
\vspace{-2cm}
\caption{{\it From left to right:}IRIS SJI~1400 that shows two time-slice images while the IRIS slit scans the eruption region. The red and blue contours trace magnetic fluxes at $\pm$50~G. The Mg~{\sc ii}~k 2796~\AA, C~{\sc ii}~1335~\AA, and Si~{\sc iv}~1394~\AA\ intensity along the IRIS slit recorded at the times of the IRIS SJI images and at the location seen at the projected IRIS slit on the SJIs. The C~{\sc ii} and Si~{\sc iv} are shown in a reversed colour table. The location between the black horizontal lines is where the mini-filament is recorded, while the part along the slit between the white horizontal lines is the reconnection site area.} 
\label{fig4}
\end{figure*}

 The IRIS data provide a unique opportunity to investigate for the first time an erupting MF and associated phenomena in simultaneous imaging and spectroscopic observations. We analysed three of the recorded spectral lines, the chromospheric optically thick Mg~{\sc ii}~k 2796.35~\AA\ (hereafter Mg~{\sc ii} k~2796), and C~{\sc ii}~1334.53~\AA\ (hereafter C~{\sc ii}~1335) lines, and the transition region optically thin Si~{\sc iv}~1393.75~\AA\ (hereafter Si~{\sc iv}~1394) line. The \mg\ and \c\ have a complex formation and in the present event, the emission seen along the slit is even more intricate given the fact that it comes from plasmas which constitute an erupting filament, a reconnection site, and micro-flare ribbons \citep{2020A&A...643A..19M}. The central depression in the Mg~{\sc ii} line is referred to as k$_3$, while the two peaks on of both sides are known as k$_{2v}$ and k$_{2r}$. For simplicity, we will refer to these peaks as blue and red wings. 
 
Figure~\ref{fig5}, right panel, displays the Doppler velocity image in the \si\ line obtained from a single Gaussian fit. It demonstrates a very clear image of a bi-directional flow that is centred above the polarity inversion line of the cancelling bipole. We obtained the reference wavelength from a line profile averaged over the entire dataset that includes two raster scans. It should be noted that after the slit crosses the region indicated with the white arrow the dynamic event was already reaching its final moments which is why the blue- and red-shifted are no longer detectable.

The IRIS slit first enters the eruption site seen as a bright elongated feature in the IRIS images after 11:12~UT (shown in the animation in Fig.~\ref{app2}) while scanning the region from east to west. One can notice, that the slit passes over the east J-edge of the elongated bright feature in the AIA~304 and 171 images (also AIA~193 in the animation in Fig.~\ref{app1}), as early as 11:08~UT. Two minutes later this edge is obscured by the erupting mini-filament in the coronal EUV images. In the spectral data starting at 11:09:30~UT (shown in the animation in Fig.~\ref{app2}) the mini-filament shows red-shifted emission first (around Solar Y 111\arcsec) at just a few \kms\ in the \si\ and \c\ lines. Emission decreases in the red wing of \mg\ which probably indicates the expanding MF rather than falling down cool MF material. Unfortunately, the spatial resolution does not permit us to determine this with more certainty.

In Fig.~\ref{fig4} we display the emission in the three spectral lines at two times during the eruption, at 11:13~UT and 11:16~UT. During the 11:13~UT exposure, the slit crosses the east edge of the eruption and displays very complex emission and absorption line-shift patterns in the \mg\ line. One should note here that generally filaments' cores have very low temperatures and thus their plasma is partially-ionised. Together with their high density, this makes them optically thick to certain wavelengths including \mg\ and \c. The \c\ and the \si\ emission comes from the filament (prominence)-corona transition region (PCTR) \citep[for details on PCTR see][]{2014LRSP...11....1P}. The MF is seen in absorption in the \mg\ line as its cool material absorbs the incident radiation which also happens in the \c\ line. At 11:13~UT in the \mg\ one can distinguish a decrease of the emission (an absorption feature) in the blue  wing of the \mg\ line along the spectrometer slit (see the two black horizontal lines that mark the location along the slit where the filament is recorded) which indicates that cool material is lifting up (Doppler shifts of up to approximately 50~\kms). The same line shifts are also recorded in the \c\ and \si\ lines. Erupting large filaments have rarely been recorded in spectral data but in the occasions, this has happened, they display with very large Doppler velocities of up to 300~\kms\ \citep[e.g.][]{2015A&A...575A..39M,2016A&A...588A...6G}. Such an event has been simultaneously registered in unique spectral SUMER (Solar Ultraviolet Measurements of Emitted Radiation on board the Solar Dynamics Observatory, SoHO) and CDS (Coronal Diagnostic Spectrometer/SoHO), and TRACE (Transition Region and Coronal Explorer) imaging observations. For details see figures 4, 5, 6 and 7 in \cite{2015A&A...575A..39M}. A similar filament eruption was recorded in data from EIS (Extreme-ultraviolet imaging spectrometer) on board Hinode \cite[see figure 5 in][]{2016A&A...588A...6G}. 

During the 11:16~UT exposure, the blue-shifted emission associated with the filament is still present in the wing of the \mg\ data (intensity decrease in the wing, marked with the upper horizontal black lines in Fig.~\ref{fig4}). One can also see a very bright blue-shifted wing in the \si\ line of more than 100~\kms\ and weaker in the \c\ line. One should be cautious though that at this slit position the emission in the spectral data especially in the transition region lines is possibly composed of emission coming along the line-of-sight from both, the mini-filament and the bright elongated feature. Red-shifted emission of up to 20~\kms\ is also clearly seen lower along the slit (see the lower two horizontal black lines) and the intensity decrease in the \mg\ red-wing is a clear indication that it is related to cool material, namely the mini-filament. It appears after 11:15:30~UT and it lasts during the following scan positions indicating that some or all of the filament material falls back which strongly suggest a failed filament eruption. To summarise, the red and blue shift patterns, represent the erupting filament that lifts up (blue-shifted) and expands or falls back as time progresses (red-shifted). For full details on the evolution of the erupting filament recorded simultaneously in both the imaging and the spectroscopic data please see the animation in Fig.~\ref{app2}. 

The spectroscopic data also deliver unique observations of the reconnection site. The first is that strong blue- and red-shifted emission from the reconnection site of more than 100~\kms\ (in \si) is detected in all lines at 11:12:37~UT (the animation in Fig.~\ref{app2}).  As the slit scans through the outflows  and the elongated bright feature, the blue-shifted emission is dominant and slightly offset with respect to the redshift due to the asymmetry of the structure with respect to the spectrometer slit. After the slit enters the location above the cancelling magnetic flux which is presumably the reconnection site, the blue- and red-shifted wings show similar Doppler shifts (see the animation in Fig.~\ref{app2} from 11:14:41~UT until 11:17:29~UT). This spectral emission evolution and appearance has been attributed to the so-called explosive events and will be discussed further in section~\ref{disc}. The blue-shifted wing in the Si~{\sc iv} line reaches more than 200~\kms\ at 11:14:18~UT. We should mention that the blue wing of the C~{\sc ii} line at 1335.71~\AA\ blended with C~{\sc ii}~1335.66~\AA, enters the shown spectral window when the Doppler shifts in the two wings increase above 135 km/s causing the red wing of the C~{\sc ii}~1334.53~\AA\ to overlap with the blue wings of C~{\sc ii}~1335.71 and C~{\sc ii}~1335.66~\AA\ (see the animation in Fig.~\ref{app2}). A separation of the contribution would be very challenging because the blue wing comes from two lines.

\begin{figure}[!ht]
\center
\includegraphics[scale=0.16]{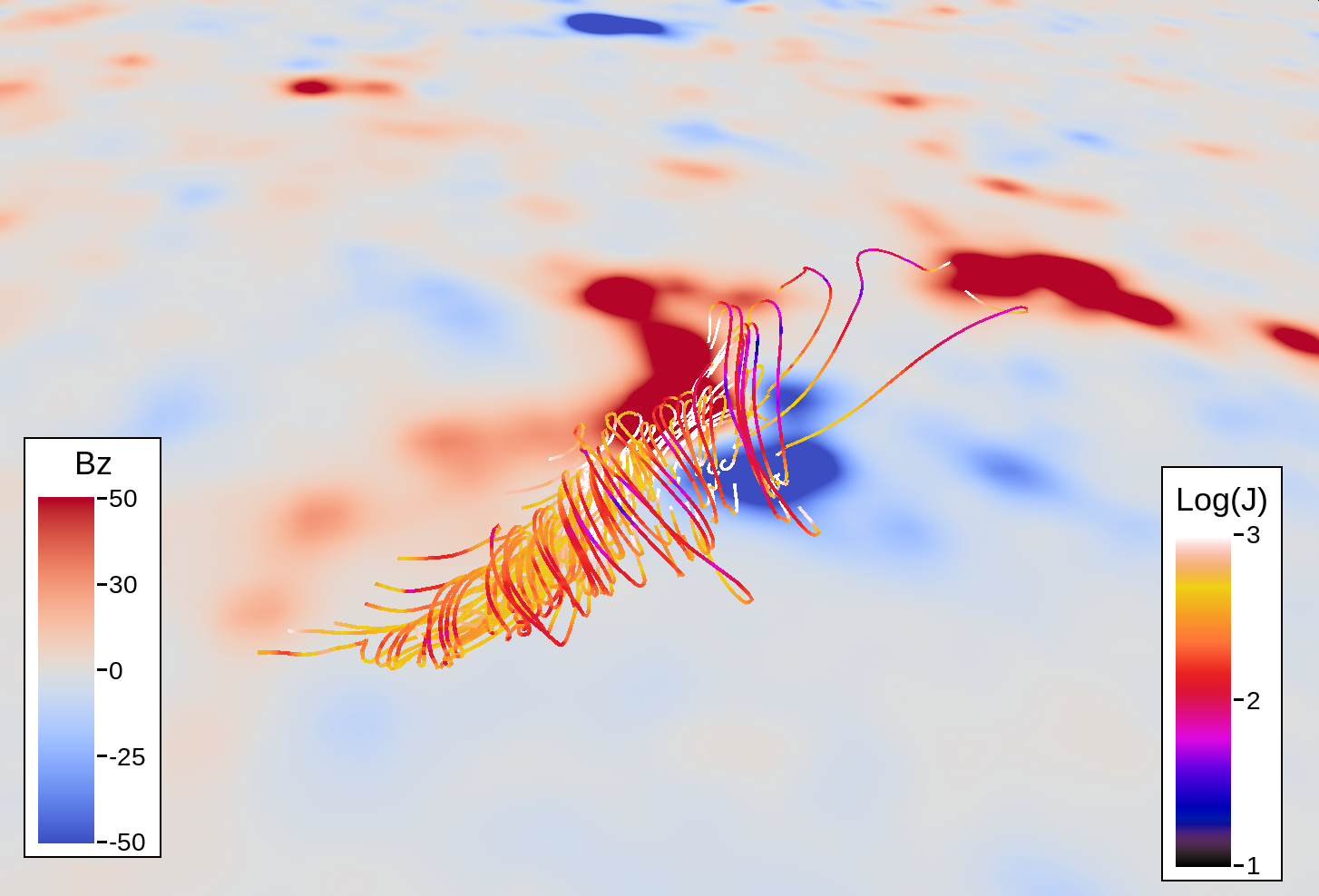}\\
\includegraphics[scale=0.16]{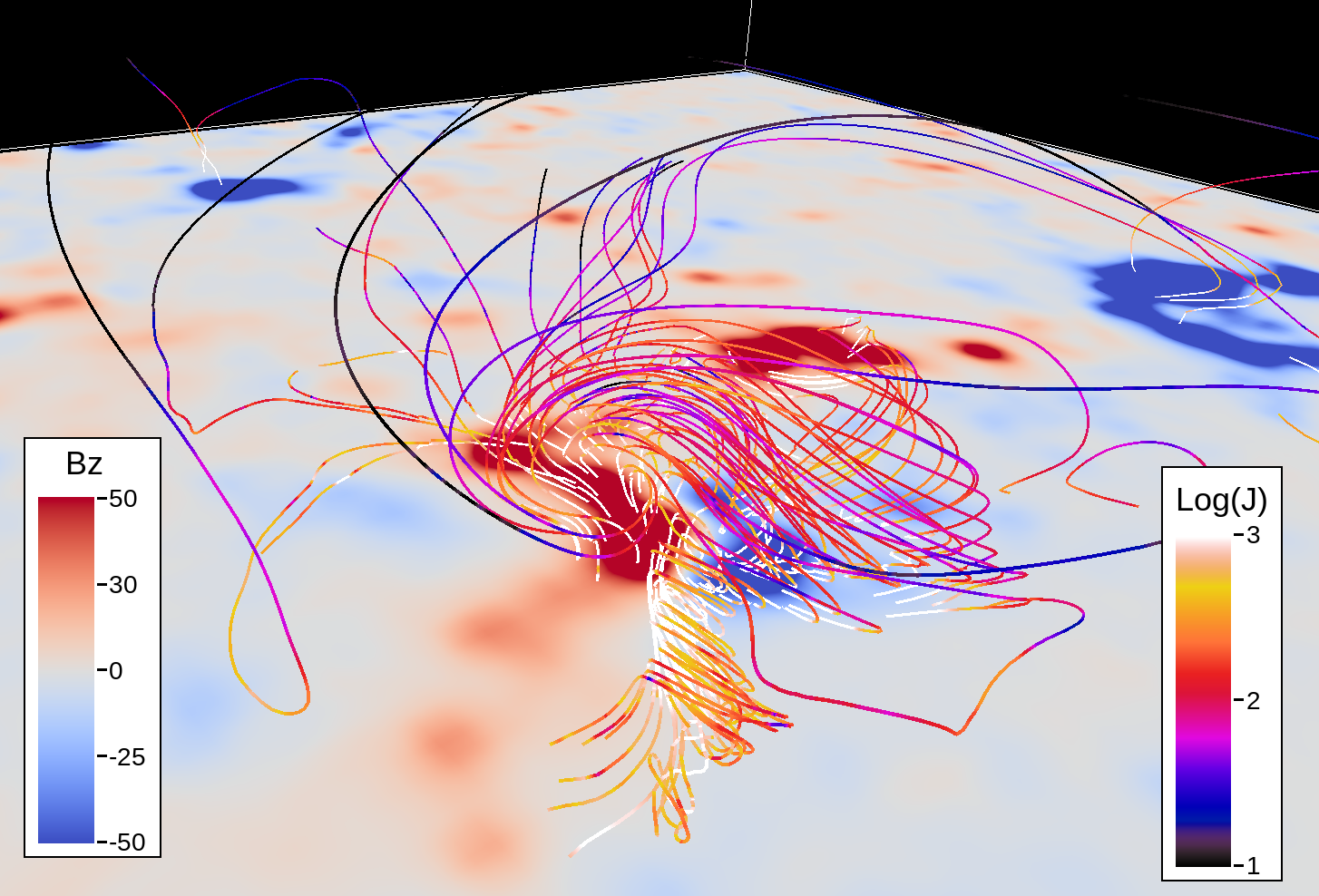}\\
\includegraphics[scale=0.16]{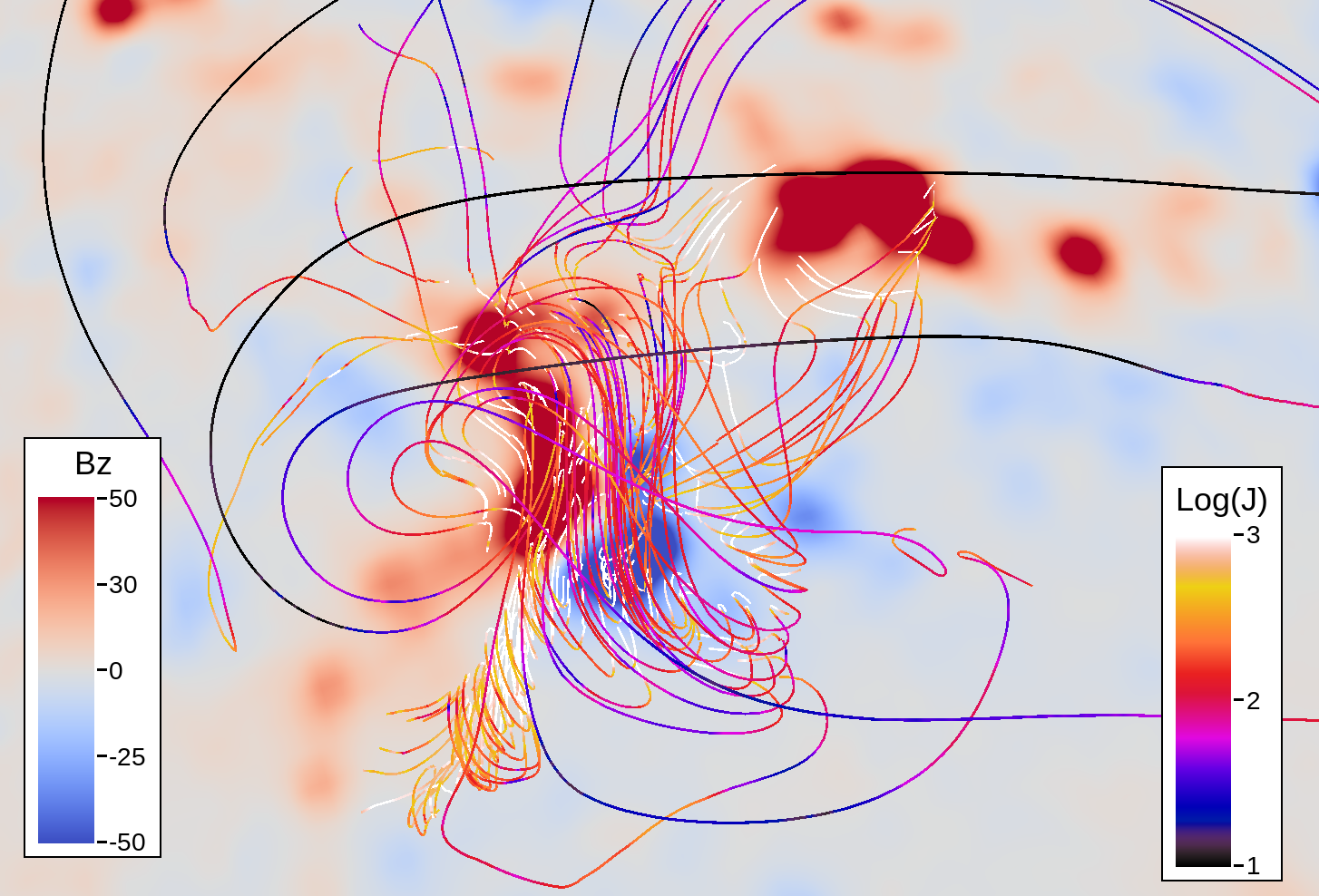}
\caption{Illustrations of the field lines above the polarity inversion line for the simulation where $\zeta=-175~$km$^2$s$^{-1}$ at 10:58 UT on 2017 April 5. The field lines are shown at low (top) and high (middle and bottom) heights above the PIL. The colour coding represents the strength of the electric current, defined as ${\bf J} = \nabla \times {\bf B}$ along the field lines as given in the right-hand side (Log(J) ranging from 1 to 3). The background image represents the photospheric magnetogram where red represents positive flux and blue negative flux  in the range $\pm$50~G, left-hand sub-panel. }
\label{fig6}
\end{figure}

\subsection{NLFFF modelling}
\label{er_mod}
To study the non-potential evolution of the CBP coronal magnetic field we simulated a continuous time series of NLFFFs.
The NLFFFs are produced using the technique described in the papers of \citet{2011ApJ...729...97M} and 
\citet{2014ApJ...782...71G}. In this technique, a time series of HMI line-of-sight magnetograms are directly applied to simulate
the boundary driving at the solar photosphere. The initial coronal magnetic field is chosen to be a potential magnetic field deduced from
the initial magnetogram in the time series. The simulations are carried out in an open and periodic computational box of
512$^3$ grid points, where the resolution in the vertical direction is set to be twice that used in the horizontal directions.
To specify the normal magnetic field component on the lower boundary, a 256$^2$ HMI pixel region centred on the region of 
interest, is cut out of the HMI magnetograms and then interpolated to 512$^2$ while conserving flux. This increased resolution 
is used in the simulations so that small-scale structures in the low solar corona can be resolved, along with minimising the effect 
of numerical diffusion.
Due to the application of open-top boundary conditions, no correction for flux balancing is applied and a significant amount of the 
photospheric flux reaches the top boundary and is defined as open. Following this, as the time sequence of normal component 
magnetograms is applied, the
coronal magnetic field responds to these driving motions by evolving through a continuous series of quasi-static NLFFF using the 
magneto-frictional relaxation technique. Full details of the numerical implementation and the equations solved can be found in the cited 
papers along with Paper~II and III.

First, we produced a series of NLFFF simulations following the description given in \cite{2014ApJ...782...71G} where the only injection of
energy and non-potentiality into the solar corona arises from the time evolution of the magnetic footpoints deduced from the 
normal component magnetograms. The simulations start at 01:58~UT on 2017 April 5 and run until 11:35~UT on the same day.
The resulting 3D non-potential magnetic field is then analysed at the location of the CBP around the time of the observed
eruption (11:00 -- 11:10~UT). These initial simulations failed to produce any strongly sheared or highly non-potential magnetic structures
at the location or time of the CBP or MF. In contrast, the field lines maintained a simple nested loop structure normal to the PIL. This structure
closely represented what would be expected from a potential field. Further investigation showed that the lack of strongly sheared fields
was a consequence of the simple flow patterns of the magnetic elements around the CBP.
These flow patterns involved only a simple converging motion of the positive and negative polarities over the time period considered, 
without any shear flows. Such motions cannot inject any axial field long the PIL to explain the CBP or MF structure, but rather just 
result in field lines that shrink in length and height while remaining normal to the PIL. Due to this, we conclude that for the
present example the observed horizontal motions deduced from the normal component magnetograms, 
when acting on an initial potential field cannot inject enough
non-potentiality into the coronal field to reproduce the observed phenomena.To investigate the validity of the initial potential field AIA
171, 193 and 211~\AA\ observations taken at 01:58~UT were studied. The visible, but faint coronal loops were then compared to the field lines of the
initial potential field. The comparison produced a good agreement between the observations and the potential field model indicating that there was no pre-existing shear in the field at this time. As such the discrepancy between the observation and model found at 11:00~UT was not due to the initial condition. At this point it is important to note that the simulated magnetic features are not fully resolved as their scale is close to the resolution of the HMI magnetograms. Due to this we may speculate that there 
may be unresolved motions such as shearing motions or vortical flows that lead to a further injection of non-potentiality into the field. 
Such additional injection is necessary to explain the highly non-potential field configurations of the CBP and MF seen in the observations.

\begin{figure}[!ht]
\vspace{-15.5cm}
\hspace{-4.5cm}
\includegraphics[scale=0.47]{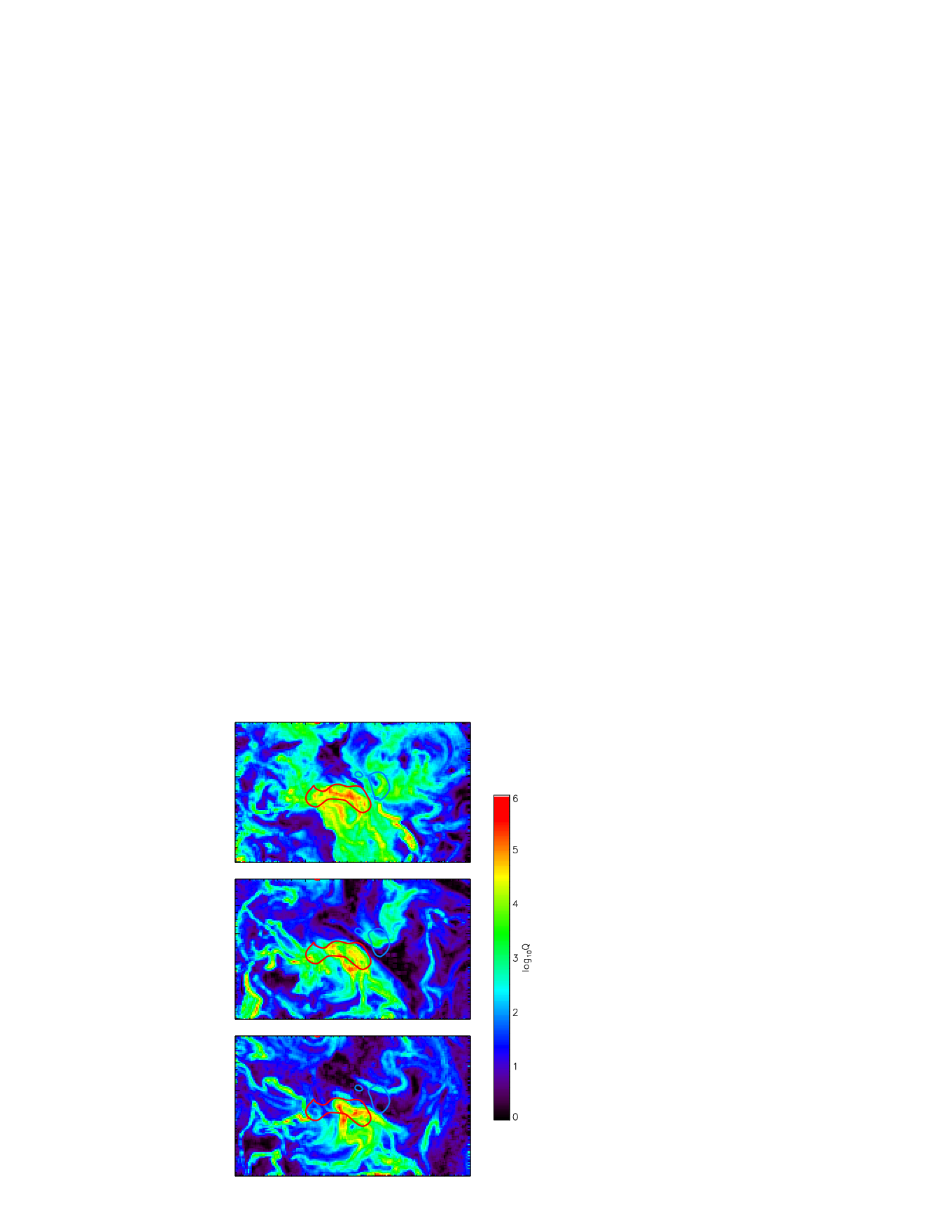}
\vspace{-1cm}
\caption{ Three frames showing the spatial distribution of the derived log$_{10}Q$ values in different horizontal planes above the magnetogram surface. The top panel corresponds to a plane at a height of 110~km, the middle at 1105~km, and the bottom at 2764~km.  The blue and red contours denote the negative and positive photospheric magnetic field, respectively at values of $B_z$=$\pm$50~Gauss. The images are smoothed with a boxcar  of width 3$\times$3 px$^2$.}
\label{fig7}
\end{figure}

As the basic approach described above only uses the normal component magnetograms and since it lacks the ability to reproduce the
main features of the observations, we considered a number of speculative experiments where additional non-potentiality was injected
into the coronal field. To carry this out the same potential field initial condition as used in the previous experiments at 01:58~UT on 2017 April 5
was taken.  Next, while applying the same evolution of the normal field component at the photosphere an additional axial field
is injected into the simulation at the lower boundary by including the term,
\begin{equation}
\frac{\partial {\bf{A}}_h}{\partial t} = - \nabla_h \left( \zeta B_z \right)
\end{equation}
where ${\bf{A}}_h = (A_x,A_y)$ represents the horizontal components of the magnetic vector potential at the photospheric boundary ($z=0$),
$B_z$ the normal magnetic field component and $\zeta$ the rate of injection of the horizontal magnetic field or twist component, 
where it has the dimensions of diffusivity. The addition of this term does not alter the evolution of $B_z$ but injects a horizontal magnetic field
component given by,
\begin{equation}
\frac{\partial {\bf{B}}_h}{\partial t} = - \nabla_z \times \nabla_h \left( \zeta B_z \right)
\end{equation}
at the lower boundary which then propagates upwards along the magnetic field lines through the ideal ${\bf{v}} \times {\bf{B}}$ term
in the induction equation. This term only acts where there is a horizontal spatial variation of the quantity $\zeta B_z$ and injects
a horizontal field at locations where the gradient is strong such as at polarity inversion lines.
The sign or twist of the horizontal field that is injected depends on the parameter $\zeta$
where a positive (negative) value of $\zeta$ leads to the injection of negative (positive) magnetic helicity. 
This term is mathematically equivalent to that used in \cite{2014ApJ...784..164M} to simulate the spatially averaged
long term effects of the helicity condensation process of \cite{2013ApJ...772...72A} without having to resolve individual 
convective cells. As the magnetic fragments in the present simulations are close to the resolution of HMI it is possible
that there are unresolved vortical motions below the presently resolved scales. These may provide an additional source of
non-potentiality not captured by the previously applied horizontal motions. Due to this we now carry out speculative simulations 
where an additional non-potential field is injected during the time evolution of the field.

Using the formulation described above along with the time sequence of observed normal component magnetograms, the coronal field is evolved forward from 01:58~UT on 2017 April~5 until 11:35 UT on the same day, using $\zeta= -$175~km$^2$s$^{-1}$. At various points during the evolution, the resulting field configuration was investigated and compared to the observations. The general tendency was that
the longer the injection the more increased shear was found along the PIL. The three images in Fig.~\ref{fig6} show examples of the resulting field lines around the PIL at 10:58 UT on 2017 April 5. The top image represents the field structure of the extended lower part of the PIL. The middle and bottom frames in Fig. \ref{fig6} include higher-lying field lines that show a sheared arcade system overlying the PIL. 
The colour coding of the field lines represents the strength of the current, defined as ${\bf J} =\nabla \times {\bf B}$.  This indicates that the lower more strongly sheared field lines contain significant current  and subsequently free magnetic energy. 
The strongly  sheared field lines closely match the path of the MF, where this free magnetic energy
may be used to drive  the mini-eruption. One important aspect of the present non-potential field
is that there is no magnetic null point identified above the region along the PIL. By running several simulations and varying $\zeta$ we find this to be true no matter how much twist or axial field is injected.

 The magnetic field line traces seen in Fig.~\ref{fig6} indicate the presence of a QSL in the vicinity of the PIL between the two converging flux concentrations. This was investigated further using the data shown in Fig.~\ref{fig6}. The squashing factor Q values are determined by following the 2nd method in \citet[section 3.2]{2012A&A...541A..78P}. Figure~\ref{fig7} shows the $log10$ of the derived Q value over the relevant part of the magnetogram. To better see the location of the high-Q layers, magnetic field contours of the  normal photospheric component of the magnetic field at $\pm$50~Gauss values are also shown. The Q visualisation reveals a complicated spatial distribution of the Q-value in the lower atmosphere. It shows the existence of different complex-shaped high-value Q-layers on either side of the region containing the sheared field lines in Fig.~\ref{fig6}. The different frames in Fig.~\ref{fig7} show how the Q structure and spatial location changes with height above the photosphere and how the different structures shift relative to the underlying magnetic field concentrations. From the calculations, it is seen that the Q-layers for the lowest height are concentrated well within the two converging polarity regions. As expected with increased height in the atmosphere they are located closer to the projected PIL. It is also found that the Q-layers extend in the southeast direction relative to the closest approach of the two flux concentrations.

\section{Discussion}
\label{disc}

 Small-scale eruptions also known as mini-CMEs in the quiet Sun are now known to be omnipresent with as many as 1400 per day over the whole Sun \citep{2009A&A...495..319I}. No less than 870 of them are believed to be related to CBPs, Paper~I. These small-scale eruptions eject both cool and hot plasma into the upper atmosphere and are expected to have a certain (maybe even significant) contribution to the energy and mass balance of the solar atmosphere. Here we used unique imaging and spectral co-observations to investigate the nature of one of these dynamic phenomena and more specifically the physical processes involved in them. We report IRIS, both spectroscopic and imaging, and AIA observations of a mini-filament eruption and a micro-flare. The appearance of an elongated bright feature whose J-shaped edges wrap around the footpoints of the erupting mini-filament strongly suggests reconnection in a quasi-separatrix layer (QSL) formed through the convergence of small-scale magnetic polarities. These polarities observed in the HMI magnetograms strongly resemble their large counterparts in active regions. 

 These observations also reveal important details on the properties of these eruptions including a relation of the reconnection driven outflows seen in the IRIS imaging and spectroscopic data to ``explosive events''. As mentioned in Section~\ref{er_sp}, the spectral-line profiles detected while crossing the eruption site, are usually attributed to the so-called ``explosive events''. Explosive events were discovered by \citet{1983ApJ...272..329B} in C~{\sc iv} 1548.21~\AA\ in spectra taken with the high-resolution telescope and spectrometer (HRTS) and were first called turbulent events. As the term ``turbulent'' did not appear to describe the observed spectral-line widths, they were then named ``explosive events'' (EEs) \citep{1984ApJ...281..870D, 1986ApJ...310..456D, 1989SoPh..123...41D}. These EEs display strong blue and red-shifted emission, where often the blue emission is stronger than the red, and the line profiles can be either Gaussian or non-Gaussian. They are short-lived (60--350~s) 
but often appear in bursts continuing for as long as 30 minutes in regions undergoing magnetic cancellation \citep[e.g.][]{1994AdSpR..14d..13D, 1998ApJ...504L.123C}. Their sizes determined along a spectrometer slit are 3\arcsec--5\arcsec. They display Doppler shifts of up to 200~\kms. \citet{1991JGR....96.9399D} first suggested
that this should be the spectroscopic signature of high-velocity bi-directional plasma ejections during magnetic reconnection. \citet{1998ApJ...504L.123C} associated EEs with mixed magnetic polarity or found them at the border of regions with a large concentration of magnetic
flux. The authors reported that out of 163 EEs, 103 occurred above areas of magnetic-flux cancellation. The EEs were also registered in data taken by the SUMER spectrometer and IRIS \citep[e.g.][]{1997Natur.386..811I, 2014ApJ...797...88H}. More details on the background of EEs can be found in \citet{2014ApJ...797...88H}. From the EE description above it became quickly evident that the spectroscopic observations presented here display an EE. The EEs indeed represent bi-directional flows from a reconnection site, but these outflows do not solely contribute to this emission.  The IRIS and AIA imaging information indicates the contribution from an erupting mini-filament and eventually chromospheric evaporation of gentle type have contributed to the EEs, i.e. non-Gaussian spectral line profiles.

 The evolution of the mini-eruption described here involving an elongated bright feature with J-shaped edges seen in the EUV (for details see section~3.1) follows very closely the flux cancellation model for CMEs of \cite{2010ApJ...708..314A}. This model was inspired by the observations of the evolution of cancelling bipolar magnetic flux in active regions. The cancellation observed here (see animation in Fig.~\ref{app1}) without any doubt is very similar to active region flux cancellation but on a much smaller scale. 
Numerous models have demonstrated the formation of  sheared magnetic flux due to magnetic flux cancellation along the polarity inversion line and/or the shearing of arcade field lines \citep[e.g.][and references therein]{2010SSRv..151..333M}. It has long been suggested that CBPs represent scaled-down active regions. During the late stages of the CBPs lifetime, the magnetic bipoles that sustain these phenomena, are observed to cancel in the majority of the cases \cite[for details see][]{2016ApJ...818....9M}. Eruptions in active regions also typically occur in the late stage of the AR lifetime, the so-called decay phase \citep[e.g.][]{2002A&A...382..650D,2009ApJ...700L..83G}. The mini-filaments' formation and evolution reported in \citet{2018A&A...619A..55M} and the event in the present study also strongly support this similarity. Because of the small scale of the observed feature and the small number of pixels in which it is resolved in the observations some details cannot be observed or derived from the modelling which is discussed in detail in section~3.3. Surprisingly similar to the model, the photospheric magnetic flux shows a decrease of 24\% during the three consecutive eruptions. 

The narrow elongated feature with J-shaped edges (sigmoid-like) with a width of only 0.34\arcsec\ ($\sim$250~km) as seen in the IRIS SJ 1400 images presumably represents the newly reconnected loops. Almost a pixel (IRIS) size brightening is identified as the micro-flare initial reconnection site. The brightening quickly spreads bi-directionally along the elongated feature. An investigation of the imaging information from all AIA channels analysed here indicates emission from plasmas heated to up to several million degrees (Fe~{\sc xviii}, 7~MK formation temperature). The highest (\aahhh) temperature emission is localised in smaller areas than the rest of the high-temperature emission (\aah\ and \aahh). 
 Both edges of the elongated structure curl into a J-shaped feature (hook-like) creating the appearance of a sigmoid at the end of which plasma ouflows are observed and are best seen in the IRIS SJIs (see Fig.~\ref{app1}). This coincides with the lift-off of the erupting mini-filament with the hooks' located at both ends of the mini-filament. 
\citet{2012ApJ...750...15S} compared the geometrical and topological properties of sigmoidal active region CME involving a filament using an NLFFF model and high-resolution magneto-hydrodynamic (MHD) simulations. The study established that in both the MHD simulations and the observed magnetic field evolution derived from the NLFFF model, flux cancellation plays a fundamental role in building the magnetic flux rope. The main QSLs were found to wrap around the edges of the flux rope. The investigation also showed the existence of hyperbolic flux tubes (HFTs) \citep{2007ApJ...660..863T} in the MHD models in vertical cross-sections of the computed squashing factor, with the main HFT, located under the  sheared flux
rope in both models. This HFT was identified as the most probable site for reconnection. From their torus instability analysis \citet{2012ApJ...750...15S} concluded that the eruption of the flux rope was caused by the combination of the flux rope expanding in a torus instability domain together with magnetic reconnection in the HFT. 
 The mini-filament slowly rises for approximately 6~min (see Table~1), before a brusque liftoff. This may have been caused by the interplay of the magnetic reconnection and a torus \citep[e.g.][]{2010ApJ...718.1388D,2006PhRvL..96y5002K} or kink instability \citep[][]{2004A&A...413L..23K,2005ApJ...630L..97T}. Due to the small scale of the erupting filament in comparison to large eruptions \citep[e.g.][and reference therein]{2012A&A...540A.127K}, it is impossible to say in the present case whether one or the other instability has played a role in the destabilisation and eruption. 
The scenario of QSL reconnection involving a filament is described by \citet[][and references therein]{2014ApJ...788...60J} with a cartoon illustration shown in Figure~7. Visualisation of the vertical current layer with clearly distinguishable hooks is shown in Figure~11 of \citet{2013ApJ...779..129K}.  The signature of the micro-flare observed as intense localised in a few pixels brightening is observed below the rising mini-filament. As shown in Paper~III, this  implies that high energy release, presumably from magnetic reconnection,  triggered by the rising mini-filament results in the heating of the chromospheric plasma  to high temperatures causing the appearance of (micro-)flare ribbons  (unresolved here due to resolution limitations) beneath the rising filament as shown in detail in Paper III.

The squashing factor Q investigation based on the advected and vortex imposed NLFFF modeling, shows the presence of a number of QSLs in the region around the PIL between the two converging flux concentrations (see Fig.~\ref{fig7}). These QSLs nicely align with the sheared flux tube identified in Fig.~\ref{fig6}. Despite the somewhat `artificial' approach in obtaining this field line configuration, it clearly indicates that the sheared field structure is embedded in a QSL that defines a possible location where magnetic reconnection between a rising flux rope and the ambient magnetic field may take place, creating the general action discussed from the combined observations. Imposing an artificial twist to the magnetic field to build up the flux-rope magnetic structure was required  because of the limitations of the available observations. In the future, we hope that ground-based observations from, e.g., the Visible Spector-Polarimeter on Daniel K. Inouye Solar Telescope (DKIST) at resolution as low as 50~km may help resolve the fine details of the photospheric and coronal magnetic field evolution so it may be better modelled.

Some of the detected upflows (blue-shifted emission) in the chromospheric and transition-region lines may be related to chromospheric evaporation. Due to the small scale of the event, it is impossible to precisely separate the emission from the reconnection site, chromospheric evaporation, and erupting mini-filament. Recently, \citet{2020A&A...643A..27F} have demonstrated that particle beams in the quiet Sun and especially from micro-flares like the one studied here can cause chromospheric heating. This would lead to chromospheric evaporation, so although speculative, the suggestion for the presence of chromospheric evaporation is in our view very plausible. If the data have been taken in a sit-and-stare mode maybe we could be able to say more with greater certainty like in \citet{2016A&A...588A...6G} during a large filament eruption. The other challenge is the size of the event or rather the resolution of the present instrumentation. The spectral data analysis showed, however, a lack of a distinctive red-shifted emission in the transition-region and chromospheric lines as commonly observed during explosive chromospheric evaporation. This hints at a possible presence of gentle chromospheric evaporation interpretation of the detected lower velocity upflows. It has been estimated by \cite{1985ApJ...289..434F} that when the energy flux during flares is lower than $\sim$10$^{10}$ ergs cm$^{-2}$ s$^{-1}$, the heating of the chromosphere would lead to gentle evaporation upflows in a wide range of temperatures and at a speed of several tens \kms\ with no accompanied downflows. Typically gentle evaporation is reported during the decay phase of solar flares and relates to plasma heated by thermal conduction during this phase \cite[e.g.][]{1988ApJ...329..456Z,2005A&A...430..679B}. \cite{2006ApJ...642L.169M} report to have observed gentle evaporation during the impulsive phase of a C9.1 flare driven by non-thermal electrons. Figure~4 in \cite{2006ApJ...642L.169M} demonstrates the difference in the Doppler velocity in spectral lines with different temperatures with blue-shift in the case of gentle evaporation and red-shift in the case of an explosive one in transition region lines. IRIS does not have strong enough coronal lines to account for the line shifts at coronal temperatures.
 A forthcoming study by \citet{hannah2022} reports the detection of hard X-ray emission in the 2.5--4~keV (stronger) and 4--6~keV (much weaker) by the Nuclear Spectroscopic Telescope Array (NuSTAR) in an event similar to the one described here, confirming the presence of micro-flares in small-scale eruptions in the quiet Sun which would trigger gentle chromospheric evaporation. We hope to detect a few more of these events in IRIS data so we can deliver further details especially on the detection of chromospheric evaporation during small-scale eruptions in the quiet Sun.

\section{Summary and Conclusions}
 This paper reports on the investigation of an eruption from a coronal bright point associated with a mini-filament eruption and cancelling magnetic fluxes of opposite polarities (Fig.~\ref{fig1}). Remarkably all the phenomena are recorded in both the imaging and spectroscopic data of the Interface Region Imaging Spectrograph. We identify an intense small-scale brightening as a micro-flare, followed by the mini-eruption associated with the ejection of the CBP loops and mini-filament plasmas (Fig.~\ref{fig2} and animation in Fig~\ref{app1}). The mini-filament eruption is seen in the spectroscopic IRIS data as spectral-line shifts (stronger blue- and weaker red-shifts) of the Mg~{\sc ii}, C~{\sc ii}, and Si~{\sc iv} lines  (Figs~\ref{fig5} and \ref{fig4}). An elongated bright feature (located above the polarity inversion line) only visible in the IRIS slit-jaw 1400~\AA\ images appears shortly before the micro-flare.  This feature appears to represent the newly reconnected loops. The micro-flare starts with an IRIS pixel size brightening and propagates bi-directionally along the elongated feature (animation in Fig~\ref{app1}).  We found strong outflows along and at the edges of the elongated feature in both the spectral and imaging IRIS data and AIA data which we believe represent the reconnection outflow. The observed red-shifted emission is seemingly related to the falling back mini-filament's plasmas. The edges of the elongated feature expand into a J-shaped feature creating the appearance of a sigmoid. 
A number of high valued Q structures are found in the region around the PIL separating the two converging flux concentrations. As coronal null points were not found above the converging and cancelling magnetic polarities, the high-value Q-regions must represent QSLs in the vicinity of the  sheared magnetic flux.

The present study uses unique IRIS observations of an eruption from a CBP and provides further observational and modelling evidence that CBPs are downscaled active regions where magnetic reconnection along QSLs, (mini-)filament eruptions, (mini-)CMEs or jets, (micro-)flaring, (gentle)-chromospheric evaporation, etc, occur. The QSL reconnection site presents in the spectroscopic data as the so-called explosive events (EEs, non-Gaussian profiles) identified by strong blue- and red-shifted emission of up to 200~\kms\ thus demonstrating the EEs indeed witness magnetic reconnection driven plasma flows.  A forthcoming study will report on the occurrence rate of mini-eruptions related to QSL reconnection in the quiet Sun and their plasma and observational properties.

\label{concl}

\begin{acknowledgements}
The authors thank very much the referee for the very important comments and suggestions. MM and TW acknowledge DFG-grant WI 3211/8-1. D.H.M. would like to acknowledge STFC for support via the Consolidated Grant SMC1/YST037. IRIS is a NASA small explorer mission developed and operated by LMSAL with mission operations executed at NASA Ames Research Centre and major contributions to downlink communications funded by ESA and the Norwegian Space Centre.
VAPOR is a product of the National Center for Atmospheric Research's Computational and Information Systems Lab. Support for VAPOR is provided by the U.S. National Science Foundation (grants $\#$ 03-25934 and 09-06379, ACI-14-40412), and by the Korea Institute of Science and Technology Information. The HMI and AIA data are provided courtesy of NASA/SDO science teams. The HMI AND AIA data have been retrieved using the Stanford University's Joint Science Operations Centre/Science Data Processing Facility. The authors thank the ISSI (Bern) for the support to the team ``Observation-Driven Modelling of Solar Phenomena''. 
\end{acknowledgements}
\bibliographystyle{aa}

\begin{thebibliography}{60}
\expandafter\ifx\csname natexlab\endcsname\relax\def\natexlab#1{#1}\fi

\bibitem[{{Antiochos}(2013)}]{2013ApJ...772...72A}
{Antiochos}, S.~K. 2013, \apj, 772, 72

\bibitem[{{Aulanier} {et~al.}(2010){Aulanier}, {T{\"o}r{\"o}k}, {D{\'e}moulin},
  \& {DeLuca}}]{2010ApJ...708..314A}
{Aulanier}, G., {T{\"o}r{\"o}k}, T., {D{\'e}moulin}, P., \& {DeLuca}, E.~E.
  2010, \apj, 708, 314

\bibitem[{{Benz}(2017)}]{2017LRSP...14....2B}
{Benz}, A.~O. 2017, Living Reviews in Solar Physics, 14, 2

\bibitem[{{Berlicki} {et~al.}(2005){Berlicki}, {Heinzel}, {Schmieder}, {Mein},
  \& {Mein}}]{2005A&A...430..679B}
{Berlicki}, A., {Heinzel}, P., {Schmieder}, B., {Mein}, P., \& {Mein}, N. 2005,
  \aap, 430, 679

\bibitem[{{Brueckner} \& {Bartoe}(1983)}]{1983ApJ...272..329B}
{Brueckner}, G.~E. \& {Bartoe}, J. D.~F. 1983, \apj, 272, 329

\bibitem[{{Chae} {et~al.}(1998){Chae}, {Wang}, {Lee}, {Goode}, \&
  {Sch{\"u}hle}}]{1998ApJ...504L.123C}
{Chae}, J., {Wang}, H., {Lee}, C.-Y., {Goode}, P.~R., \& {Sch{\"u}hle}, U.
  1998, \apjl, 504, L123

\bibitem[{{Cirigliano} {et~al.}(2004){Cirigliano}, {Vial}, \&
  {Rovira}}]{2004SoPh..223...95C}
{Cirigliano}, D., {Vial}, J.~C., \& {Rovira}, M. 2004, \solphys, 223, 95

\bibitem[{{De Pontieu} {et~al.}(2014){De Pontieu}, {Title}, {Lemen}, {Kushner},
  {Akin}, {Allard}, {Berger}, {Boerner}, {Cheung}, {Chou}, {Drake}, {Duncan},
  {Freeland}, {Heyman}, {Hoffman}, {Hurlburt}, {Lindgren}, {Mathur}, {Rehse},
  {Sabolish}, {Seguin}, {Schrijver}, {Tarbell}, {W{\"u}lser}, {Wolfson},
  {Yanari}, {Mudge}, {Nguyen-Phuc}, {Timmons}, {van Bezooijen}, {Weingrod},
  {Brookner}, {Butcher}, {Dougherty}, {Eder}, {Knagenhjelm}, {Larsen},
  {Mansir}, {Phan}, {Boyle}, {Cheimets}, {DeLuca}, {Golub}, {Gates}, {Hertz},
  {McKillop}, {Park}, {Perry}, {Podgorski}, {Reeves}, {Saar}, {Testa}, {Tian},
  {Weber}, {Dunn}, {Eccles}, {Jaeggli}, {Kankelborg}, {Mashburn}, {Pust},
  {Springer}, {Carvalho}, {Kleint}, {Marmie}, {Mazmanian}, {Pereira}, {Sawyer},
  {Strong}, {Worden}, {Carlsson}, {Hansteen}, {Leenaarts}, {Wiesmann},
  {Aloise}, {Chu}, {Bush}, {Scherrer}, {Brekke}, {Martinez-Sykora}, {Lites},
  {McIntosh}, {Uitenbroek}, {Okamoto}, {Gummin}, {Auker}, {Jerram}, {Pool}, \&
  {Waltham}}]{2014SoPh..289.2733D}
{De Pontieu}, B., {Title}, A.~M., {Lemen}, J.~R., {et~al.} 2014, \solphys, 289,
  2733

\bibitem[{{Del Zanna}(2013)}]{2013A&A...558A..73D}
{Del Zanna}, G. 2013, \aap, 558, A73

\bibitem[{{D{\'e}moulin} \& {Aulanier}(2010)}]{2010ApJ...718.1388D}
{D{\'e}moulin}, P. \& {Aulanier}, G. 2010, \apj, 718, 1388

\bibitem[{{D{\'e}moulin} {et~al.}(2002){D{\'e}moulin}, {Mandrini}, {van
  Driel-Gesztelyi}, {Thompson}, {Plunkett}, {Kov{\'a}ri}, {Aulanier}, \&
  {Young}}]{2002A&A...382..650D}
{D{\'e}moulin}, P., {Mandrini}, C.~H., {van Driel-Gesztelyi}, L., {et~al.}
  2002, \aap, 382, 650

\bibitem[{{Dere}(1994)}]{1994AdSpR..14d..13D}
{Dere}, K.~P. 1994, Advances in Space Research, 14, 13

\bibitem[{{Dere} {et~al.}(1984){Dere}, {Bartoe}, \&
  {Brueckner}}]{1984ApJ...281..870D}
{Dere}, K.~P., {Bartoe}, J. D.~F., \& {Brueckner}, G.~E. 1984, \apj, 281, 870

\bibitem[{{Dere} {et~al.}(1986){Dere}, {Bartoe}, \&
  {Brueckner}}]{1986ApJ...310..456D}
{Dere}, K.~P., {Bartoe}, J. D.~F., \& {Brueckner}, G.~E. 1986, \apj, 310, 456

\bibitem[{{Dere} {et~al.}(1989){Dere}, {Bartoe}, \&
  {Brueckner}}]{1989SoPh..123...41D}
{Dere}, K.~P., {Bartoe}, J. D.~F., \& {Brueckner}, G.~E. 1989, \solphys, 123,
  41

\bibitem[{{Dere} {et~al.}(1991){Dere}, {Bartoe}, {Brueckner}, {Ewing}, \&
  {Lund}}]{1991JGR....96.9399D}
{Dere}, K.~P., {Bartoe}, J. D.~F., {Brueckner}, G.~E., {Ewing}, J., \& {Lund},
  P. 1991, \jgr, 96, 9399

\bibitem[{{Fisher} {et~al.}(1985){Fisher}, {Canfield}, \&
  {McClymont}}]{1985ApJ...289..434F}
{Fisher}, G.~H., {Canfield}, R.~C., \& {McClymont}, A.~N. 1985, \apj, 289, 434

\bibitem[{{Fletcher} {et~al.}(2011){Fletcher}, {Dennis}, {Hudson}, {Krucker},
  {Phillips}, {Veronig}, {Battaglia}, {Bone}, {Caspi}, {Chen}, {Gallagher},
  {Grigis}, {Ji}, {Liu}, {Milligan}, \& {Temmer}}]{2011SSRv..159...19F}
{Fletcher}, L., {Dennis}, B.~R., {Hudson}, H.~S., {et~al.} 2011, \ssr, 159, 19

\bibitem[{{Frogner} {et~al.}(2020){Frogner}, {Gudiksen}, \&
  {Bakke}}]{2020A&A...643A..27F}
{Frogner}, L., {Gudiksen}, B.~V., \& {Bakke}, H. 2020, \aap, 643, A27

\bibitem[{{Galsgaard} {et~al.}(2019){Galsgaard}, {Madjarska}, {Mackay}, \&
  {Mou}}]{2019A&A...623A..78G}
{Galsgaard}, K., {Madjarska}, M.~S., {Mackay}, D.~H., \& {Mou}, C. 2019, \aap,
  623, A78

\bibitem[{{Gibb} {et~al.}(2014){Gibb}, {Mackay}, {Green}, \&
  {Meyer}}]{2014ApJ...782...71G}
{Gibb}, G.~P.~S., {Mackay}, D.~H., {Green}, L.~M., \& {Meyer}, K.~A. 2014,
  \apj, 782, 71

\bibitem[{{G{\"o}m{\"o}ry} {et~al.}(2016){G{\"o}m{\"o}ry}, {Veronig}, {Su},
  {Temmer}, \& {Thalmann}}]{2016A&A...588A...6G}
{G{\"o}m{\"o}ry}, P., {Veronig}, A.~M., {Su}, Y., {Temmer}, M., \& {Thalmann},
  J.~K. 2016, \aap, 588, A6

\bibitem[{{Green} \& {Kliem}(2009)}]{2009ApJ...700L..83G}
{Green}, L.~M. \& {Kliem}, B. 2009, \apjl, 700, L83

\bibitem[{{Green} {et~al.}(2011){Green}, {Kliem}, \&
  {Wallace}}]{2011A&A...526A...2G}
{Green}, L.~M., {Kliem}, B., \& {Wallace}, A.~J. 2011, \aap, 526, A2

\bibitem[{{Hannah} {et~al.}(2022){Hannah}, {Sterling}, {Hudson}, {Cooper},
  {Grefenstette}, {Paterson}, {Smith}, {Krucker}, {Glesener}, \&
  {White}}]{hannah2022}
{Hannah}, I., {Sterling}, A., {Hudson}, H., {et~al.} 2022, \apjl, in
  preparation

\bibitem[{{Hermans} \& {Martin}(1986)}]{1986NASCP2442..369H}
{Hermans}, L.~M. \& {Martin}, S.~F. 1986, in NASA Conference Publication, Vol.
  2442, NASA Conference Publication, 369--375

\bibitem[{{Huang} {et~al.}(2014){Huang}, {Madjarska}, {Xia}, {Doyle},
  {Galsgaard}, \& {Fu}}]{2014ApJ...797...88H}
{Huang}, Z., {Madjarska}, M.~S., {Xia}, L., {et~al.} 2014, \apj, 797, 88

\bibitem[{{Innes} {et~al.}(2009){Innes}, {Genetelli}, {Attie}, \&
  {Potts}}]{2009A&A...495..319I}
{Innes}, D.~E., {Genetelli}, A., {Attie}, R., \& {Potts}, H.~E. 2009, \aap,
  495, 319

\bibitem[{{Innes} {et~al.}(1997){Innes}, {Inhester}, {Axford}, \&
  {Wilhelm}}]{1997Natur.386..811I}
{Innes}, D.~E., {Inhester}, B., {Axford}, W.~I., \& {Wilhelm}, K. 1997, \nat,
  386, 811

\bibitem[{{Innes} {et~al.}(2010){Innes}, {McIntosh}, \&
  {Pietarila}}]{2010A&A...517L...7I}
{Innes}, D.~E., {McIntosh}, S.~W., \& {Pietarila}, A. 2010, \aap, 517, L7

\bibitem[{{Janvier} {et~al.}(2014){Janvier}, {Aulanier}, {Bommier},
  {Schmieder}, {D{\'e}moulin}, \& {Pariat}}]{2014ApJ...788...60J}
{Janvier}, M., {Aulanier}, G., {Bommier}, V., {et~al.} 2014, \apj, 788, 60

\bibitem[{{Joshi} {et~al.}(2017){Joshi}, {Kushwaha}, {Veronig}, {Dhara},
  {Shanmugaraju}, \& {Moon}}]{2017ApJ...834...42J}
{Joshi}, B., {Kushwaha}, U., {Veronig}, A.~M., {et~al.} 2017, \apj, 834, 42

\bibitem[{{Kliem} {et~al.}(2013){Kliem}, {Su}, {van Ballegooijen}, \&
  {DeLuca}}]{2013ApJ...779..129K}
{Kliem}, B., {Su}, Y.~N., {van Ballegooijen}, A.~A., \& {DeLuca}, E.~E. 2013,
  \apj, 779, 129

\bibitem[{{Kliem} {et~al.}(2004){Kliem}, {Titov}, \&
  {T{\"o}r{\"o}k}}]{2004A&A...413L..23K}
{Kliem}, B., {Titov}, V.~S., \& {T{\"o}r{\"o}k}, T. 2004, \aap, 413, L23

\bibitem[{{Kliem} \& {T{\"o}r{\"o}k}(2006)}]{2006PhRvL..96y5002K}
{Kliem}, B. \& {T{\"o}r{\"o}k}, T. 2006, \prl, 96, 255002

\bibitem[{{Koleva} {et~al.}(2012){Koleva}, {Madjarska}, {Duchlev}, {Schrijver},
  {Vial}, {Buchlin}, \& {Dechev}}]{2012A&A...540A.127K}
{Koleva}, K., {Madjarska}, M.~S., {Duchlev}, P., {et~al.} 2012, \aap, 540, A127

\bibitem[{{Leenaarts} {et~al.}(2013){Leenaarts}, {Pereira}, {Carlsson},
  {Uitenbroek}, \& {De Pontieu}}]{2013ApJ...772...90L}
{Leenaarts}, J., {Pereira}, T.~M.~D., {Carlsson}, M., {Uitenbroek}, H., \& {De
  Pontieu}, B. 2013, \apj, 772, 90

\bibitem[{{Lemen} {et~al.}(2012){Lemen}, {Title}, {Akin}, {Boerner}, {Chou},
  {Drake}, {Duncan}, {Edwards}, {Friedlaender}, {Heyman}, {Hurlburt}, {Katz},
  {Kushner}, {Levay}, {Lindgren}, {Mathur}, {McFeaters}, {Mitchell}, {Rehse},
  {Schrijver}, {Springer}, {Stern}, {Tarbell}, {Wuelser}, {Wolfson}, {Yanari},
  {Bookbinder}, {Cheimets}, {Caldwell}, {Deluca}, {Gates}, {Golub}, {Park},
  {Podgorski}, {Bush}, {Scherrer}, {Gummin}, {Smith}, {Auker}, {Jerram},
  {Pool}, {Soufli}, {Windt}, {Beardsley}, {Clapp}, {Lang}, \&
  {Waltham}}]{2012SoPh..275...17L}
{Lemen}, J.~R., {Title}, A.~M., {Akin}, D.~J., {et~al.} 2012, \solphys, 275, 17

\bibitem[{{Mackay} {et~al.}(2014){Mackay}, {DeVore}, \&
  {Antiochos}}]{2014ApJ...784..164M}
{Mackay}, D.~H., {DeVore}, C.~R., \& {Antiochos}, S.~K. 2014, \apj, 784, 164

\bibitem[{{Mackay} {et~al.}(2011){Mackay}, {Green}, \& {van
  Ballegooijen}}]{2011ApJ...729...97M}
{Mackay}, D.~H., {Green}, L.~M., \& {van Ballegooijen}, A. 2011, \apj, 729, 97

\bibitem[{{Mackay} {et~al.}(2010){Mackay}, {Karpen}, {Ballester}, {Schmieder},
  \& {Aulanier}}]{2010SSRv..151..333M}
{Mackay}, D.~H., {Karpen}, J.~T., {Ballester}, J.~L., {Schmieder}, B., \&
  {Aulanier}, G. 2010, \ssr, 151, 333

\bibitem[{{Madjarska}(2019)}]{2019LRSP...16....2M}
{Madjarska}, M.~S. 2019, Living Reviews in Solar Physics, 16, 2

\bibitem[{{Madjarska} {et~al.}(2021){Madjarska}, {Chae}, {Moreno-Insertis},
  {Hou}, {N{\'o}brega-Siverio}, {Kwak}, {Galsgaard}, \&
  {Cho}}]{2021A&A...646A.107M}
{Madjarska}, M.~S., {Chae}, J., {Moreno-Insertis}, F., {et~al.} 2021, \aap,
  646, A107

\bibitem[{{Madjarska} {et~al.}(2015){Madjarska}, {Doyle}, \&
  {Shetye}}]{2015A&A...575A..39M}
{Madjarska}, M.~S., {Doyle}, J.~G., \& {Shetye}, J. 2015, \aap, 575, A39

\bibitem[{{Madjarska} {et~al.}(2020){Madjarska}, {Galsgaard}, {Mackay},
  {Koleva}, \& {Dechev}}]{2020A&A...643A..19M}
{Madjarska}, M.~S., {Galsgaard}, K., {Mackay}, D.~H., {Koleva}, K., \&
  {Dechev}, M. 2020, \aap, 643, A19

\bibitem[{{Madjarska} {et~al.}(1999){Madjarska}, {Vial}, {Bocchialini}, \&
  {Dermendjiev}}]{1999ESASP.446..467M}
{Madjarska}, M.~S., {Vial}, J.~C., {Bocchialini}, K., \& {Dermendjiev}, V.~N.
  1999, in ESA Special Publication, Vol. 446, 8th SOHO Workshop: Plasma
  Dynamics and Diagnostics in the Solar Transition Region and Corona, ed. J.~C.
  {Vial} \& B.~{Kaldeich-Sch{\"u}}, 467

\bibitem[{{Milligan} {et~al.}(2006){Milligan}, {Gallagher}, {Mathioudakis}, \&
  {Keenan}}]{2006ApJ...642L.169M}
{Milligan}, R.~O., {Gallagher}, P.~T., {Mathioudakis}, M., \& {Keenan}, F.~P.
  2006, \apjl, 642, L169

\bibitem[{{Mou} {et~al.}(2016){Mou}, {Huang}, {Xia}, {Madjarska}, {Li}, {Fu},
  {Jiao}, \& {Hou}}]{2016ApJ...818....9M}
{Mou}, C., {Huang}, Z., {Xia}, L., {et~al.} 2016, \apj, 818, 9

\bibitem[{{Mou} {et~al.}(2018){Mou}, {Madjarska}, {Galsgaard}, \&
  {Xia}}]{2018A&A...619A..55M}
{Mou}, C., {Madjarska}, M.~S., {Galsgaard}, K., \& {Xia}, L. 2018, \aap, 619,
  A55

\bibitem[{{O'Dwyer} {et~al.}(2010){O'Dwyer}, {Del Zanna}, {Mason}, {Weber}, \&
  {Tripathi}}]{2010A&A...521A..21O}
{O'Dwyer}, B., {Del Zanna}, G., {Mason}, H.~E., {Weber}, M.~A., \& {Tripathi},
  D. 2010, \aap, 521, A21

\bibitem[{{Parenti}(2014)}]{2014LRSP...11....1P}
{Parenti}, S. 2014, Living Reviews in Solar Physics, 11, 1

\bibitem[{{Pariat} \& {D{\'e}moulin}(2012)}]{2012A&A...541A..78P}
{Pariat}, E. \& {D{\'e}moulin}, P. 2012, \aap, 541, A78

\bibitem[{{Pesnell} {et~al.}(2012){Pesnell}, {Thompson}, \&
  {Chamberlin}}]{2012SoPh..275....3P}
{Pesnell}, W.~D., {Thompson}, B.~J., \& {Chamberlin}, P.~C. 2012, \solphys,
  275, 3

\bibitem[{{Rathore} {et~al.}(2015){Rathore}, {Carlsson}, {Leenaarts}, \& {De
  Pontieu}}]{2015ApJ...811...81R}
{Rathore}, B., {Carlsson}, M., {Leenaarts}, J., \& {De Pontieu}, B. 2015, \apj,
  811, 81

\bibitem[{{Rust} \& {Kumar}(1996)}]{1996ApJ...464L.199R}
{Rust}, D.~M. \& {Kumar}, A. 1996, \apjl, 464, L199

\bibitem[{{Savcheva} {et~al.}(2012){Savcheva}, {Pariat}, {van Ballegooijen},
  {Aulanier}, \& {DeLuca}}]{2012ApJ...750...15S}
{Savcheva}, A., {Pariat}, E., {van Ballegooijen}, A., {Aulanier}, G., \&
  {DeLuca}, E. 2012, \apj, 750, 15

\bibitem[{{Scherrer} {et~al.}(2012){Scherrer}, {Schou}, {Bush}, {Kosovichev},
  {Bogart}, {Hoeksema}, {Liu}, {Duvall}, {Zhao}, {Title}, {Schrijver},
  {Tarbell}, \& {Tomczyk}}]{2012SoPh..275..207S}
{Scherrer}, P.~H., {Schou}, J., {Bush}, R.~I., {et~al.} 2012, \solphys, 275,
  207

\bibitem[{{Titov}(2007)}]{2007ApJ...660..863T}
{Titov}, V.~S. 2007, \apj, 660, 863

\bibitem[{{T{\"o}r{\"o}k} \& {Kliem}(2005)}]{2005ApJ...630L..97T}
{T{\"o}r{\"o}k}, T. \& {Kliem}, B. 2005, \apjl, 630, L97

\bibitem[{{Zarro} \& {Lemen}(1988)}]{1988ApJ...329..456Z}
{Zarro}, D.~M. \& {Lemen}, J.~R. 1988, \apj, 329, 456

\end{thebibliography}

\begin{appendix}
\section{Online material}
\label{movies}
\begin{figure*}[!ht]
\centering
\includegraphics[scale=0.26]{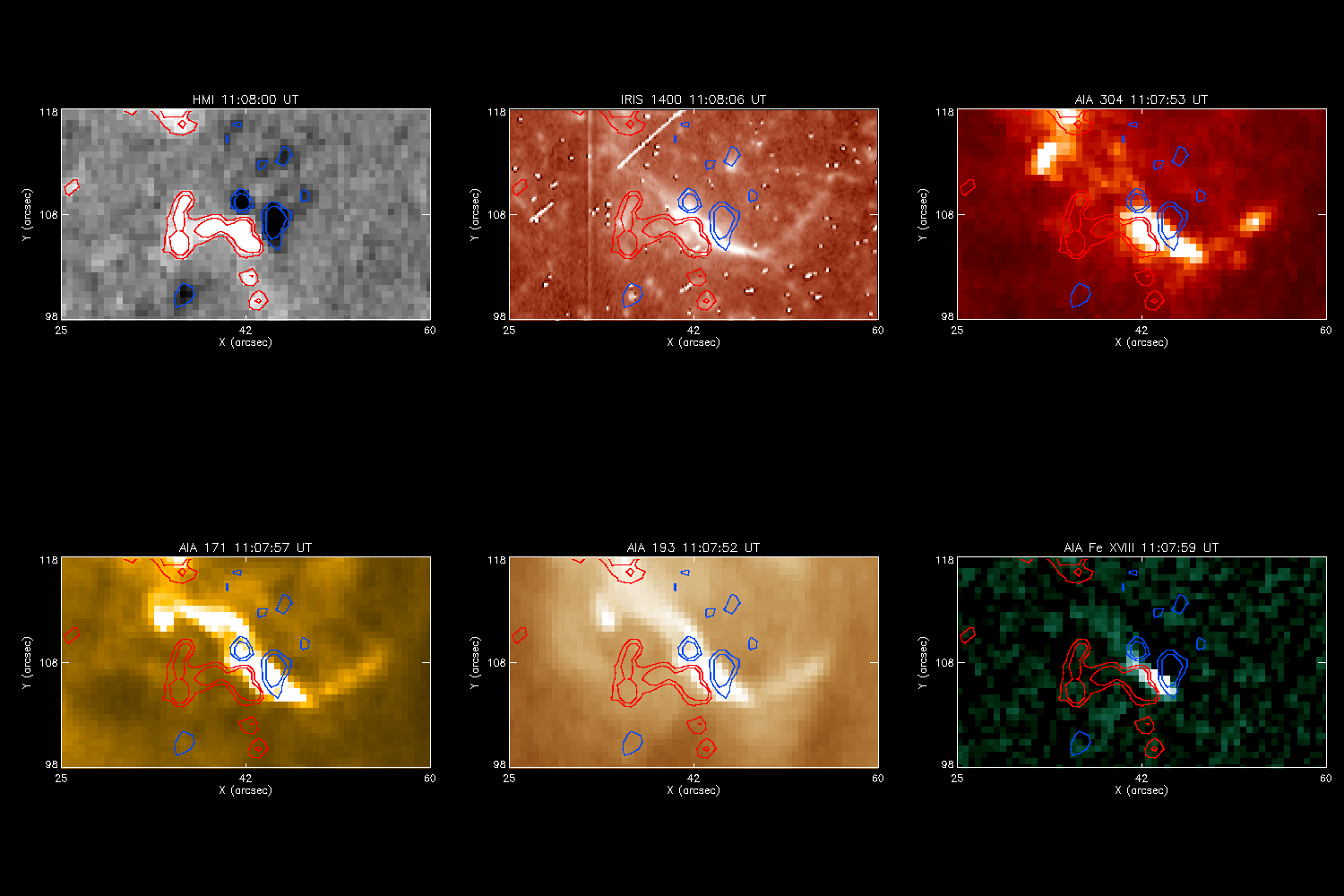}
\caption{Quasi-temporal animation. {\it From left to right, top and bottom:} HMI magnetogram, IRIS SJI 1400, \aac, 171, 193, and Fe~{\sc xviii} (see the text for details). The red and blue contours trace magnetic fluxes at $\pm$25~G and $\pm$50~G.}
\label{app1}
\end{figure*}

\begin{figure*}[!h]
\centering
\includegraphics[scale=0.26]{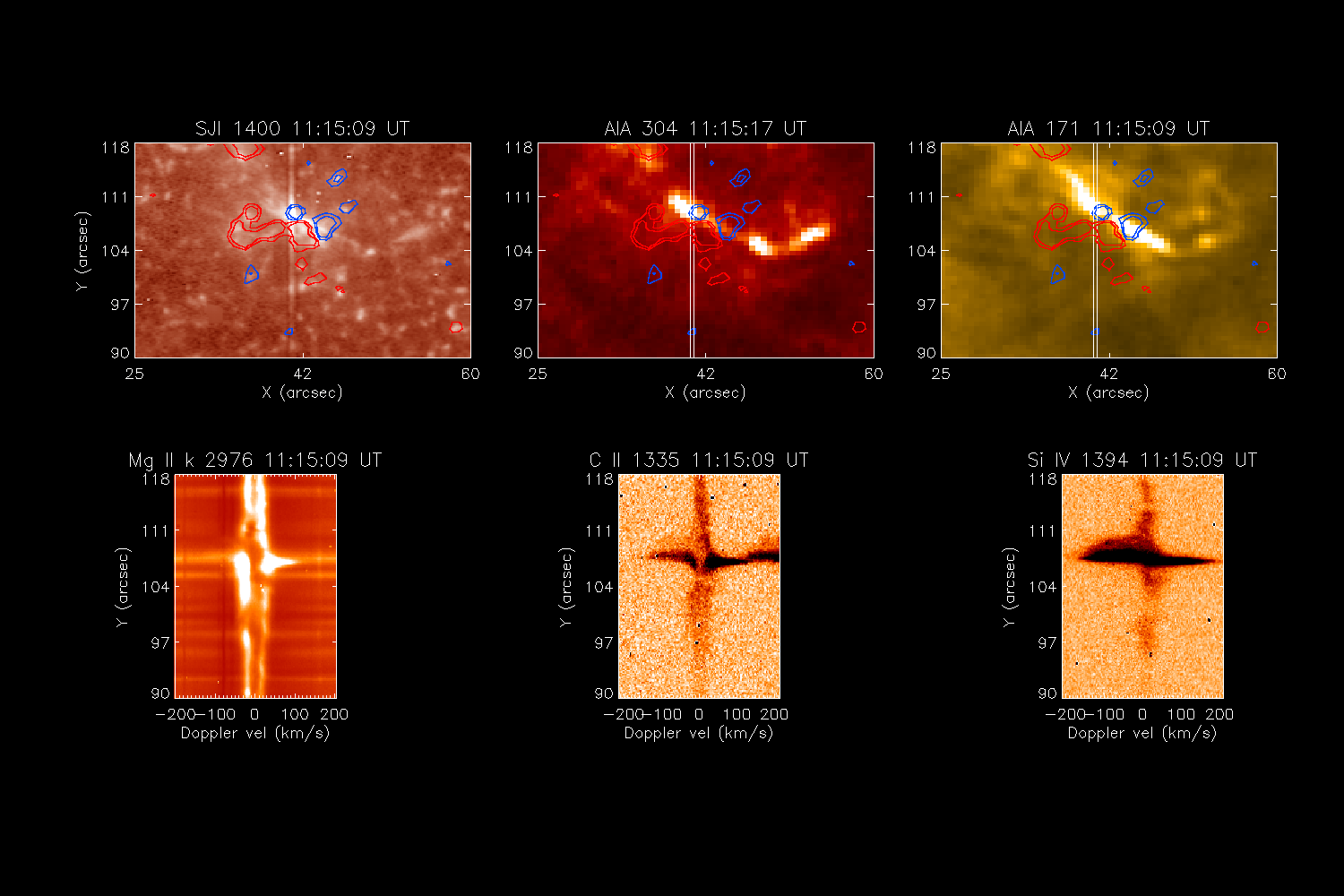}
\caption{Co-temporal animation of IRIS SJI~1400 and spectral slit images: {\it Top:} IRIS SJI~1400 in original colour table and rainbow-plus-black colour table to highlight the dynamically evolving eruption. The red and blue contours trace magnetic fluxes at $\pm$25~G and $\pm$50~G. {\it Bottom:} \mg, \c, and \si\ intensity images along the IRIS slit. The position of the slit is visible as a vertical strip on the SJI 1400 images in the left panels.} 
\label{app2}
\end{figure*}
\end{appendix}

\end{document}